\documentclass[iop,numberedappendix]{emulateapj}
\pdfoutput=1
\usepackage{graphicx}
\usepackage{rotating}
\usepackage{threeparttablex}
\usepackage{verbatim}
\usepackage{amsmath, amsthm, amssymb}
\usepackage{subfigure,captcont}

\slugcomment{{\sc Accepted to ApJ:} 2016 February 15}
\shorttitle{Resolving GMCs in NGC 300}
\shortauthors{Faesi et al.}

\newcommand{\kms}{\rm{km}~\rm{s}$^{-1}$}
\newcommand{\HII}{\ion{H}{2}}
\newcommand{\CO}{$^{12}$\rm{CO}}
\newcommand{\Msun}{\ensuremath{M_{\odot}}}

\begin{document}

\title{Resolving Giant Molecular Clouds in NGC 300: \\ A First Look with the Submillimeter Array}
\author{Christopher M. Faesi\altaffilmark{1,2}, Charles J. Lada\altaffilmark{1}, Jan Forbrich\altaffilmark{3,1}}
\altaffiltext{1}{Harvard-Smithsonian Center for Astrophysics, 60 Garden Street, Cambridge, MA 02138}
\altaffiltext{2}{NSF Graduate Research Fellow}
\altaffiltext{3}{University of Vienna}

\begin{abstract}
We present the first high angular resolution study of giant molecular clouds (GMCs) in the nearby spiral galaxy NGC~300, based on observations from the Submillimeter Array (SMA). We target eleven 500~pc-sized regions of active star formation within the galaxy in the {\CO}(J=2-1) line at 40~pc spatial and 1~{\kms} spectral resolution and identify 45 individual GMCs. We characterize the physical properties of these GMCs, and find that they are similar to GMCs in the disks of the Milky Way and other nearby spiral galaxies. For example, the GMC mass spectrum in our sample has a slope of $1.80 \pm 0.07$. Twelve clouds are spatially resolved by our observations, of which ten have virial mass estimates that agree to within a factor of two with mass estimates derived directly from {\CO} integrated intensity, suggesting that the majority of these GMCs are bound. The resolved clouds show consistency with Larson's fundamental relations between size, linewidth, and mass observed in the Milky Way. We find that the linewidth scales with the size as $\Delta V \propto R^{0.52\pm0.20}$, and the median surface density in the subsample is $54~\Msun$~pc$^{-2}$. We detect $^{13}$CO in four GMCs and find a mean {\CO}/$^{13}$CO flux ratio of 6.2. Our interferometric observations recover between 30\% and 100\% of the integrated intensity from the APEX single dish {\CO} observations of \cite{Faesi:2014ib}, suggesting the presence of low-mass GMCs and/or diffuse gas below our sensitivity limit. The fraction of APEX emission recovered increases with the SMA total intensity as well as with the star formation rate.

\end{abstract}


\section{Introduction}
\label{sec:intro}

Giant Molecular Clouds (GMCs) are the localized, cold, high-density condensations that are found within the interstellar medium (ISM), and in which gas is converted into stars. The majority of these structures are located in the disks of spiral galaxies such as the Milky Way, and thus galaxy disks represent the primary mode of star formation in the local universe \citep[e.g.,][]{2004MNRAS.351.1151B} and out to redshift $z \sim 2$, where the cosmic star formation rate density peaks \citep[e.g.,][]{2011ApJ...739L..40R}. To understand the fundamental process of star formation, and in particular the initial conditions which GMCs represent, it is thus of key importance to investigate the physical properties of GMCs. The Milky Way presents a fruitful yet limiting laboratory for such investigations. The high spatial resolution achieved in local studies has led to detailed knowledge of the structure and physical properties of the nearest GMCs from core ($<0.1$~pc) to cloud (10-100~pc) scales \citep[e.g.,][and references therein]{Heyer:2015ee}. However, our vantage point within the Milky Way disk leads to difficulty in separating clouds along a given line-of-sight, particularly when observing toward the inner Galaxy. Furthermore, derived physical distances are notoriously uncertain within the Galaxy, leading to large uncertainties in fundamental GMC physical properties that depend on distance, such as size and mass.

\begin{deluxetable*}{l c c c c c c c c c l}[!tbp]
\centering
\tabletypesize{\footnotesize}
\tablecolumns{11}
\tablewidth{\linewidth}
\tablecaption{SMA observations \label{tab:obslog}}
\tablehead{
	\colhead{Date}					&
	\colhead{Source}				&
	\colhead{Array}			&
	\colhead{\#}			&
	\colhead{$D_{\rm min}$}		&
	\colhead{$D_{\rm max}$}		& 
	\colhead{$\langle \tau_0 \rangle$\tablenotemark{a}}	&
	\colhead{On-source}		&
	\colhead{Project code}			\\
	\colhead{}				&
	\colhead{}				&
	\colhead{Configuration}				&
	\colhead{Antennas}		&
	\colhead{(m)}	&
	\colhead{(m)}	&
	\colhead{(225 GHz)}			&
	\colhead{Time (min)}	&
	\colhead{}
}
\startdata
2011 Dec 6 &	DCL88-69 &	compact &	8 &	16.4 &	77.0 &	0.11 &		200.3 &	2011B-S062 \\
2013 Aug 6 &	DCL88-114 &	compact &	5 &	16.4 &	69.1 &	0.07 &		234.4 &	2013A-S062 \\
2013 Aug 9 &	DCL88-137C &	compact &	5 &	16.4 &	69.1 &	0.16 &		297.2 &	2013A-S062 \\
2013 Aug 12 &	DCL88-137C &	compact	&	5 &	16.4 &	69.1 &	0.20 &		297.7 &	2013A-S062 \\
2013 Dec 24 &	DCL88-137B &	compact &	6 &	16.4 &	69.1 &	0.07 &		209.2 &	2013B-S074 \\
2013 Dec 29 &	DCL88-52 &	compact &	6 &	16.4 &	69.1 &	0.11 &		171.6 &	2013B-S074 \\
2014 Jan 8 &	DCL88-79 &	compact &	6 &	16.4 &	69.1 &	0.06 &		198.8 &	2013B-S074 \\
2014 Sept 21 &	DCL88-41 &	compact-north & 8 &	16.4 &	139.2 &	0.12 &		297.7 &	2014A-S085 \\
2014 Oct 9 &	DCL88-23 &	compact &	7 &	16.4 &	77.0 &	0.19 &		297.2 &	2014A-S085 \\
2014 Oct 11 &	DCL88-76C &	compact &	7 &	16.4 &	77.0 &	0.23 &		296.7 &	2014A-S085 \\
2014 Oct 12 &	DCL88-119C &	compact &	7 &	16.4 &	77.0 &	0.13 &		281.9 &	2014A-S085 \\
2014 Oct 13 &	DCL88-127 &	compact &	6 &	16.4 &	68.4 &	0.09 &		312.0 &	2014A-S085 \\
2014 Nov 8 &	DCL88-76C &	compact &	7 &	16.4 &	77.0 &	0.18 &		297.7 &	2014A-S085 
\enddata
\tablenotetext{a}{Track-averaged zenith opacity at 225 GHz.}
\end{deluxetable*}

Observing populations of GMCs within nearby spiral galaxies largely alleviates the problems discussed above: all clouds are at essentially the same distance, allowing a systematic comparison of properties within a sample, and, providing the galaxy is close to face-on, line-of-sight confusion is typically absent since galaxy disks are thin. The biggest challenges in conducting extragalactic observations of GMCs are resolution (spatial and spectral) and sensitivity (to detect faint, low column density molecular gas at cloud edges that is near the minimum column density for H$_2$ self-shielding). Millimeter and sub-millimeter interferometers are thus the ideal facilities for probing extragalactic GMC populations, and numerous studies over the past two decades have conducted observations of the low-level CO rotational transitions to construct GMC catalogs in the Large Magellanic Cloud \citep[e.g.,][]{2008ApJS..178...56F}, M31 \citep[e.g.,][]{1988ApJ...328..143L,2015ApJ...798...58K}, M33 \citep[e.g.,][]{2003ApJS..149..343E,Gratier:2012km}, M51 \citep{2014ApJ...784....3C}, and other nearby ($\sim$few Mpc) spiral galaxies \citep[e.g.,][]{Rebolledo:2012ex,2013ApJ...772..107D}. Comparison with the best (if limited) catalogs in the Milky Way suggests a general similarity in GMC properties between galaxies \citep[e.g.,][and references therein]{2008ApJ...686..948B,Fukui:2010ki}. For example, GMCs in the Milky Way, M33, and M31 show similar characteristic surface densities and levels of internal turbulence, as exhibited by a similar size-linewidth relation \citep{2008ApJ...686..948B}. Measurements of the GMC mass spectrum have suggested some differences in slope between disk galaxies in the nearby universe \citep[e.g.,][]{Rosolowsky:2005gt}, but intercomparison is challenging due to differences in cloud identification methodologies, physical resolution, and selection effects between studies. Other investigations suggest a similar mass spectrum slope between nearby galaxies \citep{Fukui:2010ki}. The differences in GMC properties within an individual galaxy have been less well explored, although \cite{2014ApJ...784....3C} note that the GMC mass spectrum shape and slope in M51 GMCs differ significantly in different dynamical environments within the galaxy. Further investigation into the GMC populations of nearby face-on spiral galaxies is necessary to address fundamental questions of cloud formation, evolution, and star formation.

We have commenced a study of the nearest low-inclination spiral galaxy in the southern sky, NGC 300. To look at molecular gas on large scales, we first targeted a subset of the H~II regions identified by \cite{Deharveng:1988wh} using the APEX 12m millimeter telescope to measure CO(2-1), and analyzed publicly available \textit{Spitzer} 24$\mu$m, GALEX far-ultraviolet, and ESO 2.2m H$\alpha$ maps to derive star formation rates \citep{Faesi:2014ib}. The APEX beam has a FWHM of {27\arcsec} at 230~GHz, and thus only resolves structures of physical sizes ~$d \sim D \theta \sim 250$~pc at the 1.93~Mpc distance of NGC~300~\citep{2004AJ....128.1167G}. In order to study the molecular gas distribution on size scales characteristic of GMCs and cloud complexes (tens of pc), we have now followed up on the APEX study using interferometric observations with the Submillimeter Array (SMA). This is the first spatially resolved study of GMCs in NGC~300. We observed 11 regions from \cite{Faesi:2014ib}, each of which exhibited CO(2-1) emission in the APEX observations. These regions span a broad range of star formation rates (SFRs) and galactocentric radii. Section \ref{sec:obs} presents our observations. Section \ref{sec:clouds} describes the process by which we identified GMCs within our CO data cubes and the derivation of physical parameters. We present our GMC catalog in Section \ref{sec:results}, and discuss these results in the context of GMC populations in our own and other galaxies in Section \ref{sec:discussion}.

\section{SMA Observations}
\label{sec:obs}

\begin{figure*}[!tbp]
\includegraphics[trim={0 3in 0 3in},clip,width=\linewidth]{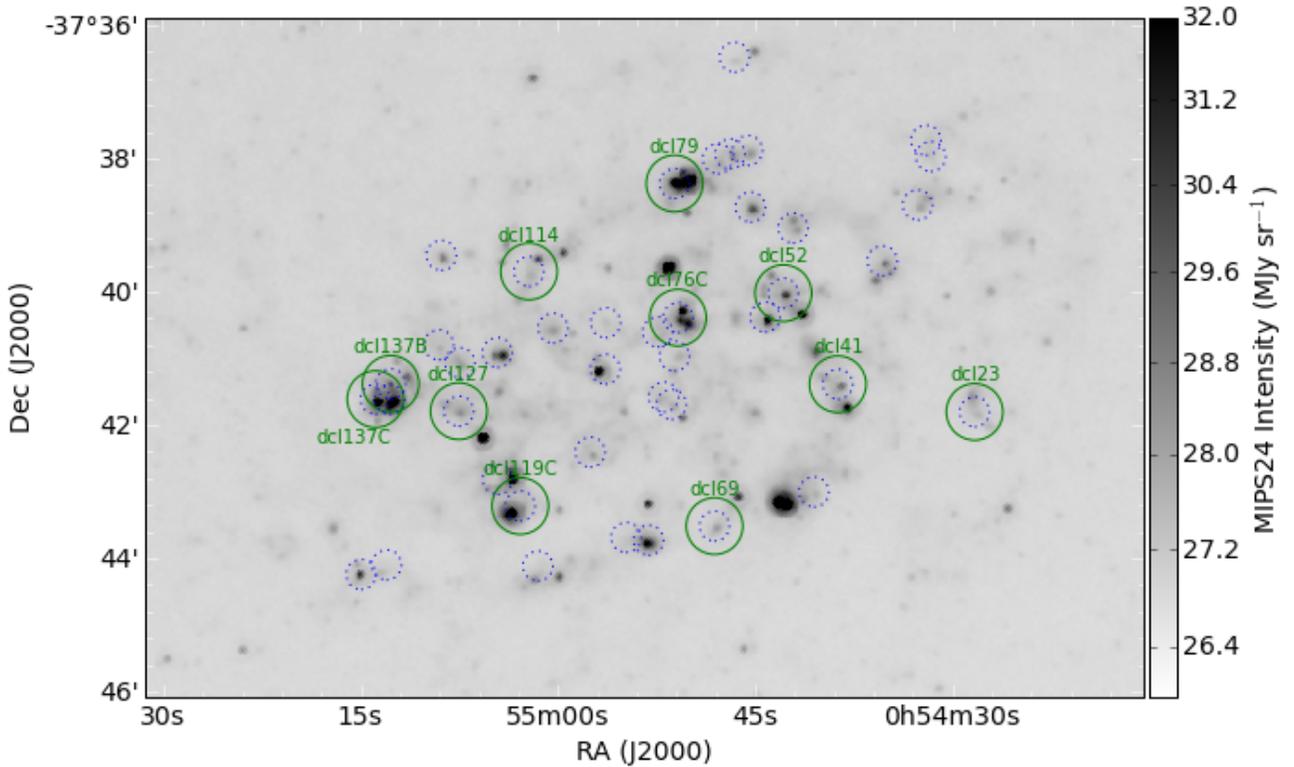}
\caption{The 11 {\HII} regions we observed with the SMA (green solid circles), which represent a subset of the full APEX sample of \cite{Faesi:2014ib} (blue dotted circles), overplotted on a \textit{Spitzer}/MIPS 24~$\mu$m image. The solid circle sizes match the SMA half-power primary beam size of $51\arcsec$ (at 230~GHz), while the dotted circle sizes match the APEX FWHM of $27\arcsec$.}
\label{fig:globalmap}
\end{figure*}

Table~\ref{tab:obslog} presents a log of our successful SMA observations of NGC 300 taken during SMA observing semesters 2011B, 2013A, 2013B, and 2014A. Our 11 targets, which represent several of the brightest CO(2-1) single dish detections from the APEX-observed {\HII} region sample of \cite{Faesi:2014ib}, were successfully observed with 13 SMA tracks over 13 nights between December 2011 and November 2014, with one source observed per track. Several additional assigned observing nights had inadequate phase coherence and/or unacceptably high atmospheric opacity, and data from these nights were not used in our analysis. Although the array nominally consists of eight antennas, we observed with between 5 and 8 antennas depending on the number in operation at the time of observation. All observations were carried out in either the compact or compact-north configuration, with minimum baselines $D_{\rm min}$ of 16.4~m and maximum baselines $D_{\rm max}$ ranging between 68 and 139~m, depending on configuration. Our average synthesized beam size was $6.1\arcsec$ by $2.6\arcsec$. $u$,$v$ coverage varied due to the night-to-night difference in observation time, array configuration, and number of operational antennas, but was typically reasonable within the range of $u$,$v$ distances spanned by the configuration, thanks to long observing tracks (3-6 hours). Figure~\ref{fig:globalmap} shows the 11 regions we observed with the SMA overlaid on a \textit{Spitzer}/Multiband Imaging Photometer \citep[MIPS;][]{2004ApJS..154...25R} 24~$\mu$m image of NGC~300. The observed regions span a range of galactocentric radii out to $\sim4$~kpc and range of SFRs from $3\times10^{-4}$ to $8\times10^{-3}~\Msun$~yr$^{-1}$ \citep{Faesi:2014ib}. The SMA primary beam is $\sim 51~\arcsec$ (FWHM) at 230~GHz.

\subsection{Setup and Data Reduction}

As a backend, we used the Application-Specific Integrated Circuit (ASIC) correlator in dual sideband mode. The lower sideband extended from 219-223 GHz and the upper sideband encompassed 229-233 GHz. Each $\sim4$~GHz sideband consisted of 48 spectral windows having 64 channels of width 1.624 MHz each. The $^{12}$CO(J=2-1) spectral line was placed in the center of the upper sideband, and the three spectral windows nearest the expected sky frequency of this line were configured for high resolution (256 channels, width 406 kHz [0.528{\kms} at the CO(2-1) rest frequency]). Our spectral setup also covered $^{13}$CO(2-1), C$^{18}$O(2-1), and SiO(4-3), but of these lines only $^{13}$CO(2-1) was detected and only in a few of the brightest {\CO} peaks in two particular regions (see \S~\ref{sec:isotopes}). Due to a tuning error, the $^{12}$CO line was placed in the lower sideband instead of the upper sideband in the DCL88-69 track, but we were still able to fully utilize this data in our analysis.

Data were reduced using the Millimeter Interferometer Reduction (\texttt{MIR}) IDL software\footnote{\url{https://www.cfa.harvard.edu/$\sim$cqi/mircook.html}} following standard procedures, which we summarize briefly here. We first flagged the raw data to remove pointing observations and bad scans. Next, we used the measured system temperatures in each antenna to scale the detected voltages into physical flux units. System temperatures were estimated during observations using the standard chopper wheel method \citep[e.g.,][]{1976ApJS...30..247U}. We calibrated for variations in phase and amplitude across the passband using observations of the quasars 3c84 and/or 3c454.3. Observations of two gain calibrators (from amongst 0137-245, 2333-237, 2258-279) were interleaved with science observations in order to correct for phase and amplitude gain variations with time. We used observations of Uranus and the online planet visibility calculator for the SMA\footnote{http://sma1.sma.hawaii.edu/planetvis.html} for final flux calibration. We adopt a conservative flux calibration uncertainty of 20\% based on comparison of our measured flux and gain calibrator visibility amplitudes with the SMA online database.

\subsection{Imaging}

After calibration, the data were exported into the \texttt{Miriad} software package~\citep{1995ASPC...77..433S} for imaging in the $^{12}$CO(2-1) line. The 16 channels on either edge of each high resolution (256 channel) spectral window were excised, and the raw visibilities were then Hanning smoothed, resulting in a final spectral resolution of 1.056~{\kms}. To account for any residual continuum emission and correct potential spectral baseline errors, we subtracted a first-order polynomial fit to the line-free channels from each integrated visibility spectrum. We imaged 100 channels centered roughly on the APEX CO line velocity, with a pixel size of 0.5\arcsec. The corresponding velocity range (105.6~{\kms}) was chosen such as to leave a factor of at least $\sim 4$ times the APEX spectral FWHM (derived from a spectral Gaussian fit to the APEX CO(2-1) spectrum) in line-free channels from which to compute the continuum RMS, and to oversample the narrowest dimension of the beam by a factor of at least four. To maximize sensitivity, we used natural weighting, resulting in synthesized beams of $6.1\arcsec$ by $2.6\arcsec$ ($57$ by $24$~pc) on average (see Table~\ref{tab:sourcelist} for synthesized beam sizes for each observation). The imaged cubes were deconvolved using the \citealp{1984A&A...137..159S} (SDI) algorithm, which minimizes the introduction of ripples (``CLEAN stripes'') in the final images. Images were cleaned to 1.5 times the average RMS (computed from emission-free channels), and then restored using the derived model for the synthesized beam. Final RMS noise estimates are listed in Table~\ref{tab:sourcelist}. The typical $1\sigma$ RMS is about 52~mJy~beam$^{-1}$ per 1.056~{\kms} channel, which corresponds to a column density of approximately $4~\Msun$~pc$^{-2}$ (using a CO(2-1)/CO(1-0) line ratio of 0.7 [\citealp{Leroy:2009di}] and a CO(1-0)-to-H$_2$ conversion factor of $2.25~\times10^{20}$~cm$^{-2}$~(K~km~s$^{-1})^{-1}$, which is the average for the regions studied here using the radial-dependent formula from \cite{Faesi:2014ib}, including a correction for helium). We assumed a linewidth of 5 {\kms} in this calculation. As a final step before analysis, we corrected each data cube for attenuation due to the SMA primary beam. The fully reduced and primary beam-corrected data cubes are publicly available for download \citep[][Dataset: \url{http://dx.doi.org/10.7910/DVN/IN7FZS}]{Faesi:2016kz}.

Sources with multiple observations were reduced separately on a per-night basis in \texttt{MIR} then combined at the beginning of the imaging procedure in \texttt{Miriad}. The DCL88-137B and DCL88-137C fields overlap significantly ($\sim 11\arcsec$ separation between pointing centers), and so we mosaicked these observations into a single data cube using a joint imaging approach in \texttt{Miriad}. Each pointing was individually imaged and then all images were linearly combined, with weights for each pixel determined by the combination of the primary beam attenuation factor and the expected noise level. Deconvolution of the mosaic was handled using the \texttt{Miriad} task \texttt{MOSSDI}, which implements an SDI clean on a mosaic. The final mosaic has an effective size over which primary beam attenuation has been corrected in a $80\arcsec$ by $40\arcsec$ field. Using a mosaic for imaging leads to significantly increased sensitivity in the large overlap area between the two observed regions as compared to imaging the regions separately.

In addition, we imaged the $^{13}$CO line (rest frequency 220.39868 GHz) similarly to $^{12}$CO. In the observing campaigns of 2011 and 2013, we used a correlator setup with high resolution (1.056 {\kms} after Hanning smoothing) channels placed at the $^{13}$CO sky frequency, and for these we used an identical imaging procedure to that described above for {\CO}. For the other observations, we achieved a final velocity resolution of 4.42~{\kms} for the $^{13}$CO images. $^{13}$CO was only detected in the brightest {\CO} clouds in DCL88-79 and DCL88-137. In DCL88-79, we also imaged C$^{18}$O, but did not detect this line. In \S~\ref{sec:isotopes} we set limits on the CO isotopologue ratios based on the measured RMS in the raw images. We also imaged the full 4~GHz continuum (excluding the {\CO} and $^{13}$CO spectral windows) using natural weighting, but did not detect any of our {\CO} sources in the 225 GHz continuum.

\begin{deluxetable*}{l c c c r r r}
\centering
\tabletypesize{\footnotesize}
\tablecolumns{7}
\tablewidth{\linewidth}
\tablecaption{Source properties \label{tab:sourcelist}}
\tablehead{
	\colhead{Region}			&
	\colhead{R.A. (J2000)}		&
	\colhead{decl. (J2000)}		&
	\colhead{rms \tablenotemark{a}}				&
	\multicolumn{3}{c}{Synthesized beam} \\
	\colhead{}									&
	\colhead{({h}\phn{m}\phn{s})}					&
	\colhead{(\phn{\arcdeg}~\phn{\arcmin}~\phn{\arcsec})}	&
	\colhead{(mJy~beam$^{-1}$)}					&
	\colhead{maj\tablenotemark{b}}							&
	\colhead{min\tablenotemark{b}}							&
	\colhead{PA\tablenotemark{d}} \\
	\colhead{} &
	\colhead{} &
	\colhead{} &
	\colhead{} &
	\colhead{($\arcsec$ / pc)} &
	\colhead{($\arcsec$ / pc)} &
	\colhead{(degree)}
}
\startdata
DCL88-23		& 00 54 28.36	& $-$37 41 48.3	& 51.3	& 5.9	/ 55	& 2.7	/ 25	& 4.4 \\
DCL88-41		& 00 54 38.75	& $-$37 41.23.5	& 45.5	& 4.9	/ 46	& 2.5	/ 23	& 12.1 \\
DCL88-52		& 00 54 42.89	& $-$37 40 01.5	& 73.9	& 6.8	/ 64	& 2.9	/ 27	& 10.3 \\
DCL88-69		& 00 54 48.11	& $-$37 43 31.3	& 31.5	& 5.5	/ 51	& 2.5	/ 23	& -13.0 \\
DCL88-76C	& 00 54 50.89	& $-$37 40 23.6	& 48.1	& 5.6	/ 52	& 2.9	/ 27	& -10.8 \\
DCL88-79		& 00 54 51.15	& $-$37 38 22.8	& 45.4	& 6.5	/ 61	& 2.5	/ 23	& 9.5	 \\
DCL88-114	& 00 55 02.20	& $-$37 39 42.0	& 51.3	& 6.5	/ 61	& 2.3	/ 22	& -17.8 \\
DCL88-119C	& 00 55 02.87	& $-$37 43 13.2	& 52.2	& 6.2	/ 58	& 2.8	/ 26	& 0.0	 \\
DCL88-127	& 00 55 07.53	& $-$37 41 47.8	& 45.5	& 6.4	/ 60	& 2.9	/ 27	& -13.1 \\
DCL88-137B\tablenotemark{c}	& 00 55 12.70	& $-$37 41 23.1	& 79.4	& 6.2	/ 58	& 2.4	/ 22	& -0.7 \\
DCL88-137C\tablenotemark{c}	& 00 55 13.86	& $-$37 41 36.9	& 79.4	& 6.2	/ 58	& 2.4	/ 22	& -0.7 
\enddata
\tablenotetext{a}{rms computed from line-free channels in dirty image.}
\tablenotetext{b}{Major and minor axis FWHM.}
\tablenotetext{c}{rms noise and synthesized beam parameters for a mosaicked image.}
\tablenotetext{d}{Measured counterclockwise from the North.}
\end{deluxetable*}

\section{Cloud Identification and Properties}
\label{sec:clouds}

We adopted a physically-motivated approach to the challenging task of identifying and characterizing GMCs from a low signal-to-noise, finite resolution data set. We employed the \texttt{CPROPS} algorithm \citep[][hereafter RL06]{Rosolowsky:2006cb}, which we briefly describe here. Note that in this paper we use the terms ``GMC'' and ``cloud'' interchangeably.

\subsection{GMC Identification}

To identify clouds, the \texttt{CPROPS} algorithm first searches the three-dimensional data cube for pairs of adjacent voxels with signal-to-noise $>4\sigma$, where $\sigma$ is the local RMS noise level per channel. It then additionally includes all other adjacent voxels with signal-to-noise $>2\sigma$ and defines each such distinct set of voxels as an ``island'' of emission. Each island is then divided into GMCs according to the following procedure. We use physical priors based on empirical knowledge of GMCs in the Milky Way to set the numerical values discussed here. First, the island is searched for local maxima, where in order to be considered, a candidate local maximum must have a brightness greater than that in all spatial pixels within a range equal to 15~pc, and also greater than that in all neighboring velocity channels within 2 {\kms}. Next, each local maximum is subjected to several critical tests: (1) the emission uniquely associated with it must be larger in area than a single synthesized beam; (2) the local maximum must be at least\footnote{The Jansky-to-Kelvin conversion factor at 230~GHz is given by $23.68 ([b_{\rm maj}/\arcsec] [b_{\rm min}/\arcsec])^{-1} \approx1.49$~Jy~beam$^{-1}$ per K on average, with exact values computed based on the synthesized beam size for each observation.} 1~K~$(15$~pc/[beam FWHM in pc])$^2$ above the merge level with the emission associated with other local maxima, where the scaling factor accounts for beam dilution, and; (3) for pairs of local maxima in a given island, the fractional change in the moments of the emission between that computed including and excluding one of the maxima must be larger than 50\%, otherwise they will not be considered distinct. The surviving local maxima are considered \textit{kernels}, and all emission uniquely associated with each is considered a ``cloud'', i.e. a GMC. The above criteria are empirically motivated based on the observed physical properties of GMCs \citep[e.g.,][see also Appendix 3 of RL06]{Solomon:1987uq}. In addition to the criteria imposed by \texttt{CPROPS}, we also require GMCs to have central pixels located inside the primary beam FWHM ($51~\arcsec$), as the noise level at this point is a factor of two higher than at the image phase center. We also require that clouds have a central velocity less than 20~{\kms} different from the APEX CO central velocity of the region from \cite{Faesi:2014ib}, as detections outside this velocity range likely represent emission physically unassociated with the galaxy at that location. This removal only affects two candidate clouds in DCL23, both of which are at vastly different velocities (+41 and +49~{\kms}) from the APEX CO velocity.

\subsection{GMC Properties}

\texttt{CPROPS} uses the voxels assigned to each cloud to compute its physical properties. The primary parameters of interest for our purposes are the GMC size $R$, FWHM linewidth $\Delta V$, and mass $M$. These quantities are computed according to the RL06 algorithm, which we summarize here. $R$ and $\Delta V$ are calculated using the intensity-weighted second moments of the one-dimensional spatial and velocity distributions, respectively, of voxels within the cloud. In general, the second moment $\sigma$ of a one-dimensional distribution is the dispersion from the mean, which reflects a measure of the width of a distribution along that dimension. The general form for the intensity-weighted second moment $\sigma_q$ of the discrete variable $q$ with intensities given by $f(q)$ is
\begin{equation}
\sigma_q = \left[\frac{\sum_i f(q_i) (q_i - \bar{q})^2}{\sum_i f(q_i)}\right]^{1/2},
\label{eqn:moment}
\end{equation}
\noindent where $f(q_i)$ is the intensity of element $i$, $\bar{q} = \sum_i f(q_i) q_i / \sum_i f(q_i)$ is the intensity-weighted first moment (i.e., the mean) of $q$, and the summation runs over all elements in the distribution. In the subsections below we explain how Equation~(\ref{eqn:moment}) is applied in \texttt{CPROPS} to derive the size and linewidth of each cloud. In the ensuing discussion, the brightness temperature $T$ replaces the general $f$ in the  equation above.

\subsubsection{GMC Sizes}

To compute the size, the \texttt{CPROPS} algorithm first rotates the $x$ and $y$ axes to align with the major and minor axes of the cloud, whose orientations are determined using principal component analysis. Next, Equation~(\ref{eqn:moment}) is applied for $q=x$ and $q=y$ to compute $\sigma_x$ and $\sigma_y$, the spatial second moments along the major and minor axis, respectively. These moments are then extrapolated to the 0~K contour, which we assume represents the outer boundary of the cloud, using a linear least-squares fit to the full set of measured contour levels above the intensity level of the cloud boundary. This extrapolation procedure corrects for the underestimation of cloud size due to the effects of finite sensitivity. Finally, the spatial beam is deconvolved from the extrapolated spatial moments to derive the direction-averaged deconvolved spatial second moment $\sigma_r$. For this calculation, we use a recently modified version of \texttt{CPROPS}\footnote{\url{https://github.com/low-sky/cprops}} kindly provided by E. Rosolowsky which generalizes the deconvolution procedure to allow for an elliptical synthesized beam at arbitrary position angle.

The size $R$ is then calculated to be $R=\eta \sigma_r$, where the numerical factor $\eta$ relates the one-dimensional RMS spatial second moment to the radius of a spherical cloud. We choose $\eta=1.91$ to match the \cite{Solomon:1987uq} empirical definition for Milky Way clouds and to facilitate direct comparison with existing Galactic and extragalactic studies~\citep[e.g.,][]{2003ApJ...599..258R,2008ApJ...686..948B,Heyer:2009ii}. This is similar to the value of $\eta \approx 2.45$ one would calculate analytically assuming a spherical cloud with power-law density distribution $\rho \propto R^{-\beta}$ and exponent $\beta=1$~(e.g., RL06). For clouds with observed sizes smaller than the synthesized beam along any axis, the GMC is considered unresolved and its size is not able to be computed.

\subsubsection{GMC Linewidths}

The velocity dispersion $\sigma_v$ is calculated using Equation~(\ref{eqn:moment}) for $q=v$. $\sigma_v$ is then extrapolated to the 0~K contour using linear least squares fitting as described above for $\sigma_r$. We then deconvolve a Gaussian approximation of the spectral response function from the extrapolated velocity dispersion $\sigma_{v,\rm{ex}}$ in order to account for the effects of finite instrumentation resolution. Specifically, $\sigma_{v,\rm{ex,dc}} = [\sigma_{v,\rm{ex}}^2 - \Delta V_{\rm chan}^2 / (2\pi)]^{1/2}$, where $\sigma_{v,\rm{ex,dc}}$ is the deconvolved velocity dispersion extrapolated to the 0~K contour, and $\Delta V_{\rm chan}$ is the width of a velocity resolution element. Following RL06, we greatly simplify the deconvolution calculation by treating the channel shape as Gaussian instead of square. The factor of $2\pi$ is the normalization such that a Gaussian-shaped channel will have the same integrated area as a square channel with width $\Delta V$. Finally, the FWHM linewidth $\Delta V$ is given by $\Delta V = \sqrt{8 \ln(2)} \sigma_{v,\rm{ex,dc}}$.

\subsubsection{GMC Masses}
\label{sec:GMCmasses}

We compute two different masses for each GMC: a luminous mass $M_{\rm lum}$, calculated by summing the CO emission over the cloud and assuming that the CO intensity is proportional to the total molecular mass; and, a dynamical or virial mass $M_{\rm VT}$, which takes as an assumption that GMCs are bound objects.

For the luminous mass, we first compute the zeroth moment of the distribution of voxels in each cloud, i.e. the CO(2-1) flux $F_{\rm CO} = \sum_i T_i \,\delta x \, \delta y \, \delta v$, where the $\delta$ factors are the voxel sizes (in {\arcsec}, {\arcsec}, and {\kms}) in the $x$, $y$, and $v$ directions, respectively, and $T_i$ is the brightness temperature at voxel $i$. $F_{\rm CO}$ thus has units of K~{\kms}~arcsec$^2$. $F_{\rm CO}$ is then extrapolated to the 0~K contour, but this time using a quadratic fit, which recovers the total integrated intensity of a model cloud more accurately than a linear fit (see RL06). $F_{\rm CO}$ is then converted into a CO(2-1) luminosity $L_{\rm CO}$ using the distance to NGC~300 (see RL06, Equation [16]), and then to molecular mass using the ``X-factor'' $\alpha_{\rm CO}$ (see below). Since NGC 300 has subsolar metallicity with an outwardly decreasing metallicity gradient, we use the customized values for the conversion factor $\alpha_{\rm CO}$ computed in \cite{Faesi:2014ib}, which accounts for the local metallicity in an azimuthally averaged sense. The luminous mass $M_{\rm lum}$ is then given by
\begin{equation}
\label{eqn:Mlum}
M_{\rm lum} = \alpha_{\rm CO} R_{21}^{-1} L_{\rm CO},
\end{equation}
\noindent where $L_{\rm CO}$ is the CO(2-1) luminosity and $R_{21}$ is the CO 2-1 to 1-0 line ratio. We take $R_{21} = 0.7$, which is the typical line ratio in the Milky Way and nearby galaxy disks \citep[e.g.,][]{1997ApJ...486..276S,Leroy:2009di}. While in reality $R_{21}$ is likely to change modestly due to region-to-region differences in excitation conditions, we do not have any independent constraint on the intrinsic line ratio in our NGC~300 sources and simply acknowledge the added uncertainty to our results.

For those GMCs we resolve (i.e., those which have observed sizes larger than the SMA synthesized beam size), we can additionally compute a virial mass, which is the gravitational mass necessary for the cloud to remain virialized (i.e., in a dynamical state in which twice the kinetic energy equals the potential energy, as described by the simplified virial theorem) given its size, velocity dispersion, and geometry. We assume spherical clouds with a truncated power law density distribution given by $\rho \propto r^{-\beta}$, where $\rho$ is the volume density and $\beta=1$ is the assumed exponent, chosen for simplicity and to match previous work \citep[e.g.,][]{Solomon:1987uq}. We also implicitly assume in using the simplified virial theorem that the energy contributed by magnetic fields and external pressure is negligible. With these assumptions, and we arrive at the following \citep[][RL06]{Solomon:1987uq}:

\begin{eqnarray}
M_{\rm VT} &=& 3~\frac{5-2\beta}{3-\beta}~\frac{\sigma_v^2 R}{G} \nonumber \\
&=& 125~\frac{5-2\beta}{3-\beta}~\left( \frac{\Delta V}{\rm{km~s}^{-1}}\right)^2 \left(\frac{R}{\rm{pc}}\right)~\Msun \nonumber \\
M_{\rm VT} &=& 189~\left( \frac{\Delta V}{\rm{km~s}^{-1}}\right)^2 \left(\frac{R}{\rm{pc}}\right)~\Msun.
\label{eqn:virialmass}
\end{eqnarray}

\subsubsection{Uncertainties in GMC properties}
\label{sec:uncertainties}

We ran \texttt{CPROPS} using the \texttt{BOOTSTRAP} flag, which calculates uncertainties on all quantities using a bootstrapping technique which we summarize here. In brief, for each cloud, the algorithm runs an additional 1000 independent times. Each time, an artificial realization of the original data set is constructed by drawing $i$ sets of $(x_i,y_i,v_i,T_i)$ one at a time from the full distribution of these quantities in the original cloud, with repetition allowed. Each moment ($\sigma_x, \sigma_y, \sigma_v$, and $F_{\rm CO}$) is then computed, and the variations in moments across all 1000 realizations is taken to be the uncertainties in those moments. The uncertainties reported in Table~\ref{tab:gmcprops} are those computed through this bootstrap method.

There are additional potential sources of uncertainty in the properties \texttt{CPROPS} reports. As a test of the reliability of extrapolated source sizes in the resolved subsample, we compared the FWHM major and minor axis lengths computed by \texttt{CPROPS} to those derived by a simple 2d Gaussian spatial fit to the CO integrated intensity map. For 2 of the 12 sources, a direct comparison was not possible, as 2d spatial fitting failed due to the integrated intensity map containing highly overlapping structures. For an additional 2 sources, the 2d Gaussian fit was successful, but the fit had to include additional pixels beyond those in the cloud being compared, and thus the source sizes derived were significantly larger. For the remaining eight sources, 2d Gaussian spatial fitting produced sources that appeared to reasonably represent the clouds \texttt{CPROPS} found. For these eight sources, 2d fitting leads to estimated FWHM major and minor axes on average 13\% larger than the extrapolated moments from \texttt{CPROPS}, although there is a range in this difference from smaller by 20\% to larger by 38\%. Position angles between these two methods always agree to within twice the reported uncertainty on position angle from the Gaussian fitting procedure. The difference in FWHM axis lengths leads to an even larger discrepancy in deconvolved source sizes. Source sizes computed from simple 2d fits are on average 50\% larger than those from \texttt{CPROPS}, and as high as 140\% larger in one case. This larger discrepancy comes from the nonlinearity of the deconvolution calculation -- the more similar the observed size and beam size, the bigger a change in deconvolved source size due to a minor change in observed size. We thus caution that in addition to the uncertainties reported by \texttt{CPROPS}, there may be a 50\% systematic uncertainty in source size.

One source in particular deserves additional mention: DCL79-2. It is the most massive GMC in our sample, and is resolved, but only marginally so (its deconvolved minor axis is 0.8~{\arcsec}). Thus its deconvolved size is estimated to be only 10~pc, which causes it to fall significantly off the scaling relations expected for GMCs (see ensuing sections). Due to the extreme nonlinearity of the deconvolution algorithm when the source size is similar to the beam size, we performed a series of tests to assess any additional uncertainties that may have led to an underestimate of this source's size. For one, we find that even small changes in the position angle could lead to drastic changes in the deconvolved size: simply rotating the cloud by only 6$^{\circ}$ leads to an increase in deconvolved size by a factor of two. Another potential uncertainty is that this source shows an extremely high central concentration, and so the intensity-weighted spatial moments appear small compared to the extent of the source out to its 2$\sigma$ contour. This may indicate that it follows a steeper relation than the $\rho \propto R^{-1}$ assumed in the calculation of $R$. We do not attempt to correct for these effects, but simply caution the reader that this source may appear to be an outlier in parameter spaces involving the size. Investigation of the density profiles of GMCs in NGC~300 would help better address this issue, but will require higher resolution observations than those presented here. The effects discussed here for DCL79-2 are minimized in the other resolved sources in our sample because their deconvolved axis lengths are sufficiently large.

\section{Results}
\label{sec:results}

\tabletypesize{\scriptsize}
\begin{deluxetable*}{ccccccccccccccccc}
\centering	
\tablecaption{NGC 300 GMC Properties \label{tab:gmcprops}}
\tablecolumns{12}
\tablewidth{500pt}
\tablehead{
\colhead{cloud ID} & \colhead{ra} & \colhead{dec} & \colhead{$v_0$} & \colhead{$d_{\rm maj}$\tablenotemark{a}} & \colhead{$d_{\rm min}$\tablenotemark{a}} & \colhead{PA\tablenotemark{b}} & \colhead{$R$} & \colhead{$\Delta V$} & \colhead{$M_{\rm lum}$} & \colhead{$M_{\rm vir}$} & \colhead{$L_{CO}$} \\
\colhead{} & \multicolumn{2}{c}{(J2000)} & \colhead{(km s$^{-1}$)} & \colhead{({\arcsec})} & \colhead{({\arcsec})} & \colhead{($^{\circ}$)} & \colhead{(pc)} & \colhead{(km s$^{-1}$)} & \colhead{($10^4 \Msun$)} & \colhead{($10^4 \Msun$)} & \colhead{(K km s$^{-1}$ pc$^2$)}
}
\startdata
DCL23-1   & 00h54m28.28s & -37d41m53.5s & 172.3 & 6.3  & 2.7 & 10  & \nodata        & $4.0 \pm 1.1$ & $7.1 \pm 1.3 $ & \nodata       & $6.44 \pm 1.16 \times 10^3$ \\
DCL23-2	  & 00h54m27.31s & -37d42m03.0s & 180.4 & 5.3  & 3.8 & 2   & \nodata        & $2.9 \pm 1.2$ & $6.5 \pm 3.0 $ & \nodata 	     & $5.92 \pm 2.72 \times 10^3$ \\
DCL41-1	  & 00h54m40.11s & -37d41m17.6s & 159.0 & 5.0  & 3.1 & 43  & \nodata        & $3.5 \pm 5.8$ & $2.2 \pm 1.2 $ & \nodata 	     & $2.54 \pm 1.40 \times 10^3$ \\
DCL41-2	  & 00h54m39.83s & -37d41m18.1s & 161.0 & 3.7  & 2.8 & -11 & \nodata        & $4.1 \pm 3.1$ & $2.9 \pm 3.4 $ & \nodata 	     & $3.36 \pm 3.90 \times 10^3$ \\
DCL41-3	  & 00h54m39.33s & -37d41m20.8s & 163.4 & 7.8  & 4.8 & 82  & $16.4\pm 7.2$  & $4.2 \pm 1.8$ & $8.7 \pm 10.0$ & $5.5 \pm 5.4$ & $1.00 \pm 1.15 \times 10^4$ \\
DCL41-4	  & 00h54m38.71s & -37d41m25.9s & 168.4 & 7.2  & 4.1 & 9   & $31.4\pm10.7$  & $8.3 \pm 2.0$ & $9.1 \pm 2.0 $ & $40.3\pm23.4$ & $1.05 \pm 0.23 \times 10^4$ \\
DCL41-5	  & 00h54m36.79s & -37d41m29.0s & 174.8 & 3.9  & 1.2 & 33  & \nodata        & $3.1 \pm 1.4$ & $1.2 \pm 1.0 $ & \nodata       & $1.38 \pm 1.16 \times 10^3$ \\
DCL41-6	  & 00h54m36.88s & -37d41m32.4s & 172.8 & 3.2  & 1.8 & 14  & \nodata        & $4.7 \pm 4.0$ & $3.2 \pm 3.2 $ & \nodata       & $3.75 \pm 3.71 \times 10^3$ \\
DCL41-7	  & 00h54m39.28s & -37d41m40.6s & 176.7 & 5.2  & 3.0 & 35  & \nodata        & $3.8 \pm 1.3$ & $4.0 \pm 2.0 $ & \nodata       & $4.63 \pm 2.27 \times 10^3$ \\
DCL52-1	  & 00h54m41.96s & -37d40m19.6s & 177.8 & 6.1  & 4.2 & 2   & \nodata        & $3.1 \pm 1.7$ & $3.1 \pm 1.5 $ & \nodata       & $3.83 \pm 1.88 \times 10^3$ \\
DCL52-2	  & 00h54m42.47s & -37d40m17.5s & 177.7 & 7.4  & 3.0 & 8   & $ 8.6\pm11.5$  & $2.8 \pm 2.1$ & $2.6 \pm 3.4 $ & $1.3 \pm 2.3$ & $3.28 \pm 4.23 \times 10^3$ \\
DCL52-3	  & 00h54m42.43s & -37d39m49.0s & 179.0 & 5.2  & 5.3 & 36  & \nodata        & $2.9 \pm 2.5$ & $2.0 \pm 1.3 $ & \nodata 	     & $2.47 \pm 1.58 \times 10^3$ \\
DCL69-1	  & 00h54m47.97s & -37d43m30.5s & 129.7 & 6.6  & 1.6 & -14 & \nodata 	    & $7.1 \pm 3.8$ & $1.4 \pm 1.0 $ & \nodata       & $1.62 \pm 1.25 \times 10^3$ \\
DCL69-2	  & 00h54m48.24s & -37d43m38.2s & 142.1 & 9.5  & 4.0 & -24 & $36.3\pm 5.4$  & $7.7 \pm 1.6$ & $15.3\pm 1.2 $ & $40.0\pm19.2$ & $1.81 \pm 0.15 \times 10^4$ \\
DCL69-3	  & 00h54m46.76s & -37d43m13.7s & 142.7 & 4.8  & 4.2 & -78 & \nodata  	    & $4.7 \pm 1.8$ & $5.4 \pm 1.7 $ & \nodata       & $6.38 \pm 1.98 \times 10^3$ \\
DCL76C-1  & 00h54m48.94s & -37d40m28.0s & 149.1 & 5.8  & 3.5 & -7  & $12.6\pm11.8$  & $6.9 \pm 1.7$ & $6.3 \pm 2.6 $ & $11.4\pm12.8$ & $8.78 \pm 3.60 \times 10^3$ \\
DCL76C-2  & 00h54m50.20s & -37d40m37.2s & 148.4 & 6.0  & 3.1 & -24 & \nodata 	    & $4.2 \pm 1.8$ & $2.7 \pm 2.1 $ & \nodata       & $3.74 \pm 2.88 \times 10^3$ \\
DCL76C-3  & 00h54m50.35s & -37d40m29.6s & 150.1 & 7.1  & 2.8 & -7  & \nodata 	    & $4.3 \pm 1.4$ & $4.3 \pm 2.3 $ & \nodata       & $5.96 \pm 3.22 \times 10^3$ \\
DCL76C-4  & 00h54m51.52s & -37d40m27.5s & 155.1 & 6.8  & 3.5 & -22 & $17.4\pm 8.2$  & $4.9 \pm 1.7$ & $3.1 \pm 1.1 $ & $7.7 \pm 6.7$ & $4.37 \pm 1.49 \times 10^3$ \\
DCL76C-5  & 00h54m50.75s & -37d40m18.7s & 156.6 & 4.2  & 2.2 & -17 & \nodata  	    & $4.1 \pm 5.0$ & $1.9 \pm 4.1 $ & \nodata       & $2.64 \pm 5.76 \times 10^3$ \\
DCL76C-6  & 00h54m50.86s & -37d40m25.8s & 159.6 & 5.6  & 2.9 & -0  & \nodata 	    & $2.0 \pm 2.1$ & $0.7 \pm 1.6 $ & \nodata       & $0.95 \pm 2.23 \times 10^3$ \\
DCL79-1   & 00h54m50.48s & -37d38m18.5s & 149.4 & 5.1  & 1.9 & -1  & \nodata 	    & $2.4 \pm 1.7$ & $0.6 \pm 0.7 $ & \nodata       & $7.78 \pm 7.94 \times 10^2$ \\
DCL79-2   & 00h54m50.06s & -37d38m20.2s & 156.0 & 8.3  & 4.2 & -14 & $10.1\pm 1.4$  & $7.4 \pm 0.6$ & $74.7\pm 7.5 $ & $10.2\pm 2.4$ & $9.06 \pm 0.91 \times 10^4$ \\
DCL79-3   & 00h54m51.35s & -37d38m22.5s & 155.0 & 7.1  & 2.8 & 1   & \nodata        & $6.1 \pm 0.9$ & $10.9\pm 1.1 $ & \nodata       & $1.32 \pm 0.13 \times 10^4$ \\
DCL79-6   & 00h54m51.81s & -37d38m38.0s & 158.3 & 6.0  & 3.1 & 3   & \nodata 	    & $6.7 \pm 2.4$ & $4.7 \pm 3.1 $ & \nodata       & $5.73 \pm 3.72 \times 10^3$ \\
DCL79-7   & 00h54m51.70s & -37d38m30.7s & 159.0 & 5.7  & 3.0 & -7  & \nodata 	    & $4.1 \pm 4.1$ & $0.5 \pm 0.8 $ & \nodata       & $6.01 \pm 9.80 \times 10^2$ \\
DCL114-2  & 00h55m02.44s & -37d39m52.2s & 111.8 & 7.8  & 2.8 & -8  & \nodata 	    & $3.5 \pm 1.5$ & $2.5 \pm 0.9 $ & \nodata       & $3.03 \pm 1.12 \times 10^3$ \\
DCL114-3  & 00h55m02.38s & -37d39m57.3s & 121.1 & 6.9  & 3.4 & -21 & $17.5\pm 4.2$  & $4.5 \pm 0.8$ & $13.0\pm 1.2 $ & $6.8 \pm 2.8$ & $1.60 \pm 0.14 \times 10^4$ \\
DCL114-5  & 00h55m01.11s & -37d39m31.2s & 127.2 & 7.1  & 3.5 & -22 & $19.5\pm14.9$  & $5.0 \pm 4.0$ & $1.9 \pm 1.2 $ & $9.0 \pm19.1$ & $2.27 \pm 1.52 \times 10^3$ \\
DCL114-6  & 00h55m01.66s & -37d39m31.4s & 126.8 & 8.7  & 2.2 & -22 & \nodata 	    & $5.0 \pm 1.7$ & $3.8 \pm 0.9 $ & \nodata       & $4.64 \pm 1.16 \times 10^3$ \\
DCL114-7  & 00h55m02.14s & -37d39m42.3s & 131.1 & 7.3  & 2.2 & -16 & \nodata 	    & $5.7 \pm 1.7$ & $3.3 \pm 1.2 $ & \nodata       & $4.08 \pm 1.47 \times 10^3$ \\
DCL119C-1 & 00h55m04.01s & -37d42m54.6s & 101.9 & 6.6  & 3.5 & -53 & \nodata 	    & $4.3 \pm 1.8$ & $3.8 \pm 2.2 $ & \nodata       & $4.60 \pm 2.67 \times 10^3$ \\
DCL119C-2 & 00h55m03.58s & -37d43m24.1s & 107.5 & 6.2  & 2.6 & -12 & \nodata 	    & $1.9 \pm 2.3$ & $1.3 \pm 2.3 $ & \nodata       & $1.55 \pm 2.77 \times 10^3$ \\
DCL119C-3 & 00h55m03.91s & -37d43m26.2s & 107.6 & 10.6 & 3.4 & 19  & \nodata 	    & $6.7 \pm 3.6$ & $5.6 \pm 1.8 $ & \nodata       & $6.76 \pm 2.16 \times 10^3$ \\
DCL119C-4 & 00h55m02.06s & -37d43m14.1s & 109.6 & 6.1  & 3.9 & 25  & \nodata 	    & $4.3 \pm 1.9$ & $1.8 \pm 1.7 $ & \nodata       & $2.19 \pm 2.01 \times 10^3$ \\
DCL119C-5 & 00h55m02.08s & -37d43m30.2s & 121.7 & 6.3  & 1.9 & -10 & \nodata 	    & $1.8 \pm 2.1$ & $1.3 \pm 1.6 $ & \nodata       & $1.53 \pm 1.97 \times 10^3$ \\
DCL127-1  & 00h55m08.95s & -37d41m43.6s & 93.5  & 5.0  & 3.2 & -40 & \nodata 	    & $3.2 \pm 1.2$ & $2.5 \pm 0.7 $ & \nodata       & $2.98 \pm 0.83 \times 10^3$ \\
DCL127-2  & 00h55m07.93s & -37d41m43.9s & 99.0  & 7.2  & 4.7 & 3   & $22.3\pm 6.2$  & $4.7 \pm 1.2$ & $7.0 \pm 0.8 $ & $9.3 \pm 6.3$ & $8.25 \pm 0.99 \times 10^3$ \\
DCL127-3  & 00h55m07.06s & -37d41m45.9s & 97.3  & 5.3  & 2.6 & -16 & \nodata  	    & $4.2 \pm 3.5$ & $0.7 \pm 2.0 $ & \nodata       & $0.83 \pm 2.34 \times 10^3$ \\
DCL127-4  & 00h55m06.75s & -37d41m36.9s & 99.2  & 6.1  & 5.4 & 80  & \nodata  	    & $2.1 \pm 0.8$ & $2.4 \pm 2.7 $ & \nodata       & $2.82 \pm 3.21 \times 10^3$ \\
DCL137-1  & 00h55m10.92s & -37d41m34.8s & 90.7  & 5.7  & 3.2 & -3  & \nodata  	    & $5.2 \pm 3.0$ & $6.0 \pm 4.9 $ & \nodata       & $6.30 \pm 5.17 \times 10^3$ \\
DCL137-2  & 00h55m13.96s & -37d41m31.5s & 96.1  & 9.1  & 4.4 & 5   & $37.7\pm 5.7$  & $6.1 \pm 0.9$ & $28.3\pm 2.3 $ & $26.3\pm 8.9$ & $2.99 \pm 0.24 \times 10^4$ \\
DCL137-3  & 00h55m12.80s & -37d41m25.4s & 88.3  & 7.2  & 2.5 & -7  & \nodata  	    & $4.3 \pm 1.6$ & $3.1 \pm 2.5 $ & \nodata       & $3.25 \pm 2.63 \times 10^3$ \\
DCL137-4  & 00h55m11.57s & -37d41m16.2s & 93.7  & 6.1  & 2.9 & 5   & \nodata 	    & $5.6 \pm 0.9$ & $22.9\pm 2.1 $ & \nodata       & $2.41 \pm 0.22 \times 10^4$ \\
DCL137-6  & 00h55m12.34s & -37d41m18.1s & 95.6  & 8.7  & 5.7 & -12 & $42.1\pm 10.5$ & $5.8 \pm 1.6$ & $12.6\pm 5.5 $ & $27.0\pm15.9$ & $1.33 \pm 0.59 \times 10^4$ \\
\enddata
\tablenotetext{a}{Extrapolated spatial FWHM diameter along the major and minor axes of the cloud (prior to deconvolution from the synthesized beam).}
\tablenotetext{b}{Measured counterclockwise from the northern declination axis, computed prior to deconvolution.}
\end{deluxetable*}

In this study we successfully resolve the CO emission seen with APEX as a single 250 pc pixel into individual Giant Molecular Clouds at $\sim40$~pc resolution. Figure~\ref{fig:COmaps} shows individual {\CO} integrated intensity maps of the 10 observed fields with ellipses representing the \texttt{CPROPS}-derived major and minor axis lengths overlaid. Each single field is approximately 470~pc across. We find that each region consists of multiple (2 to 7) discreet GMCs, with a total of 45 GMCs across all regions. Table~\ref{tab:gmcprops} presents the properties of these 45 GMCs, including the (intensity-weighted) central position in equatorial coordinates, central velocity, FWHM major and minor axis lengths, total CO(2-1) luminosity $L_{\rm CO}$, physical size $R$, linewidth $\Delta V$, mass (both luminous mass $M_{\rm lum}$ and virial mass $M_{\rm vir}$), and the associated uncertainties on these properties as computed by \texttt{CPROPS}'s bootstrap method. Twelve of these clouds have well-determined sizes from \texttt{CPROPS}; we refer to this ensemble of clouds as the ``resolved subsample'' in the ensuing discussion. The masks generated from \texttt{CPROPS} are publicly available for download along with the fully reduced SMA CO(2-1) data cubes \citep[][Dataset: \url{http://dx.doi.org/10.7910/DVN/IN7FZS}]{Faesi:2016kz}.

\begin{figure*}[!tbp]
\subfigure{\includegraphics[trim={0 2.1in 0 2.1in},clip,width=0.33\linewidth]{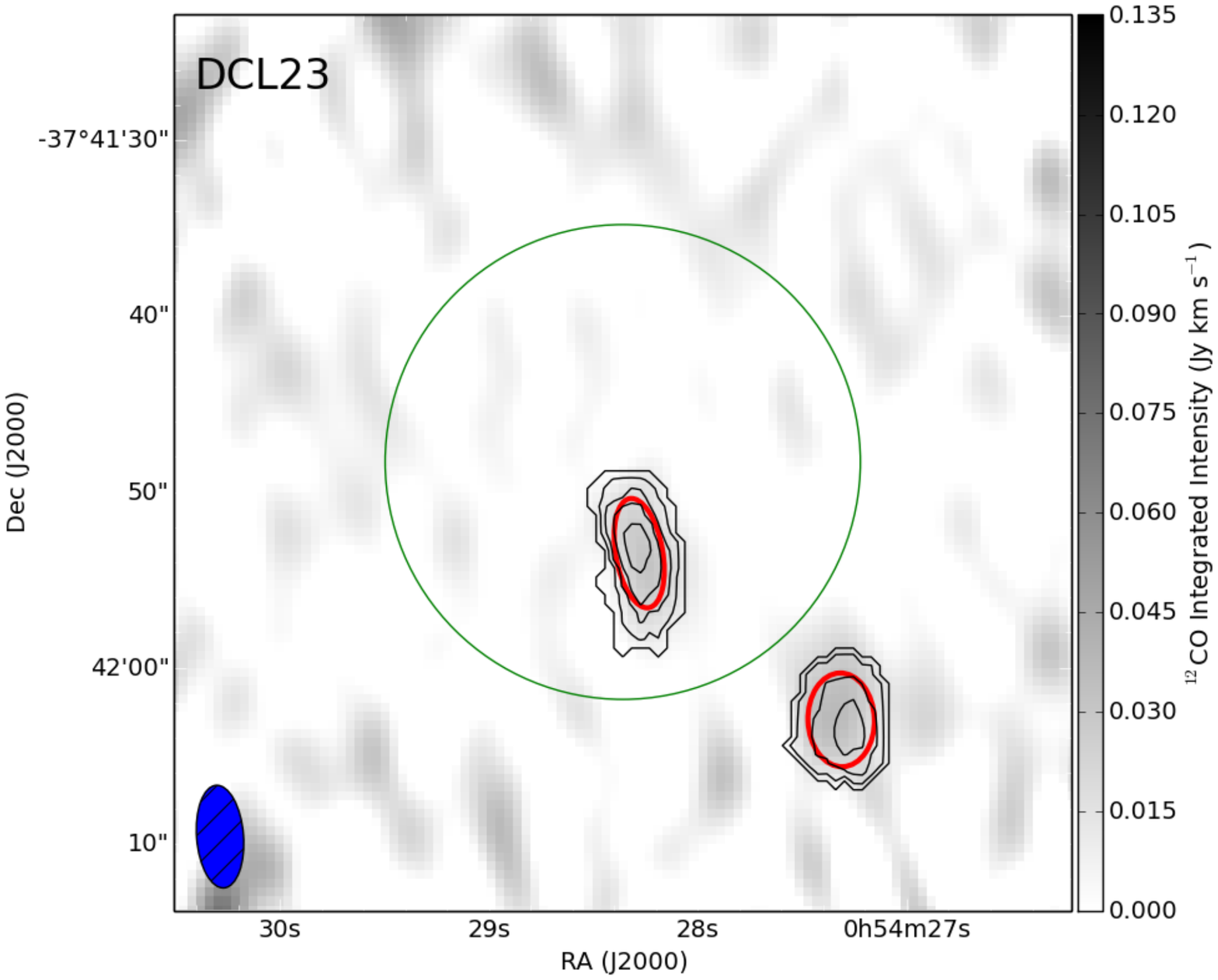}}
\subfigure{\includegraphics[trim={0 2.1in 0 2.1in},clip,width=0.33\linewidth]{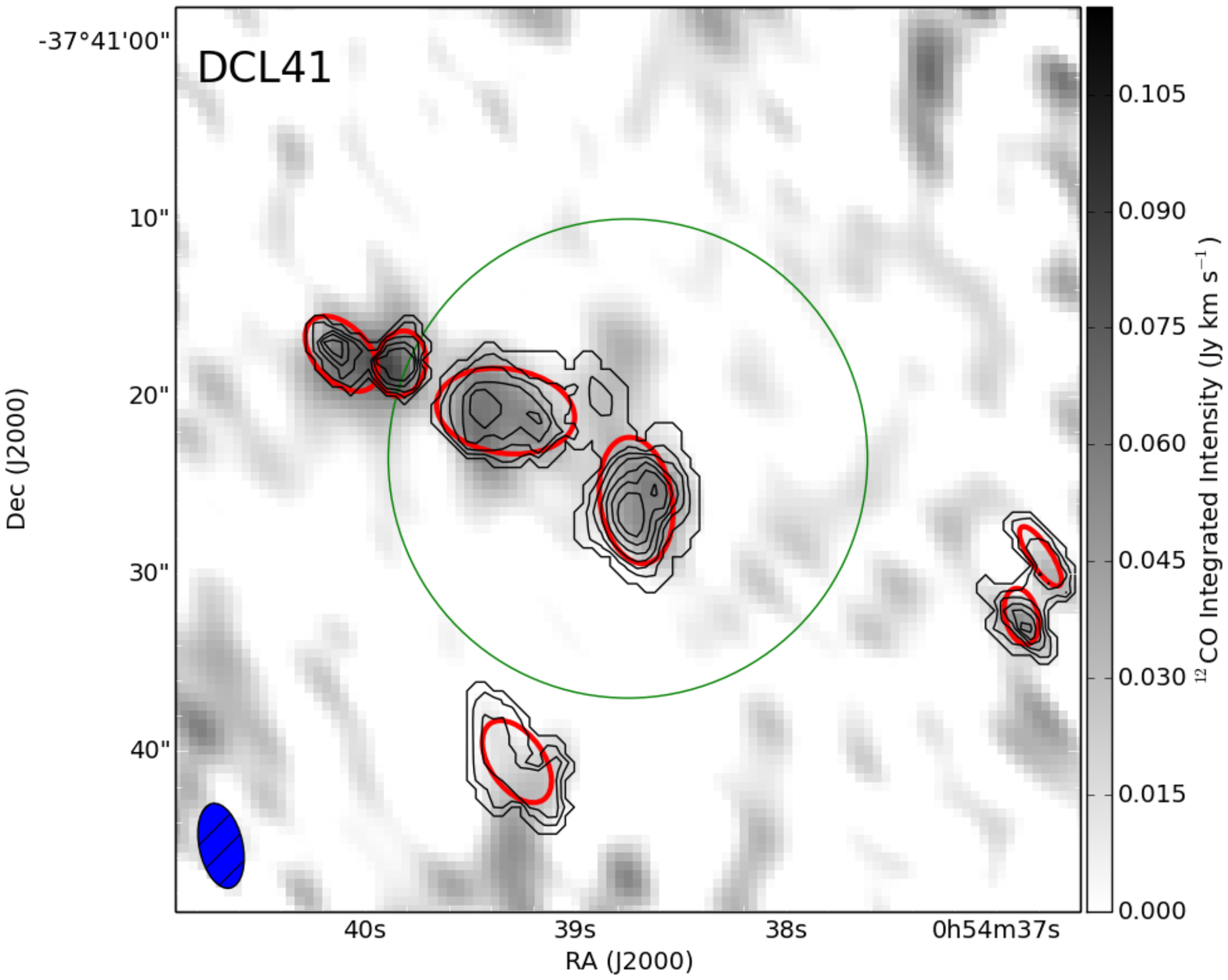}}
\subfigure{\includegraphics[trim={0 2.1in 0 2.1in},clip,width=0.33\linewidth]{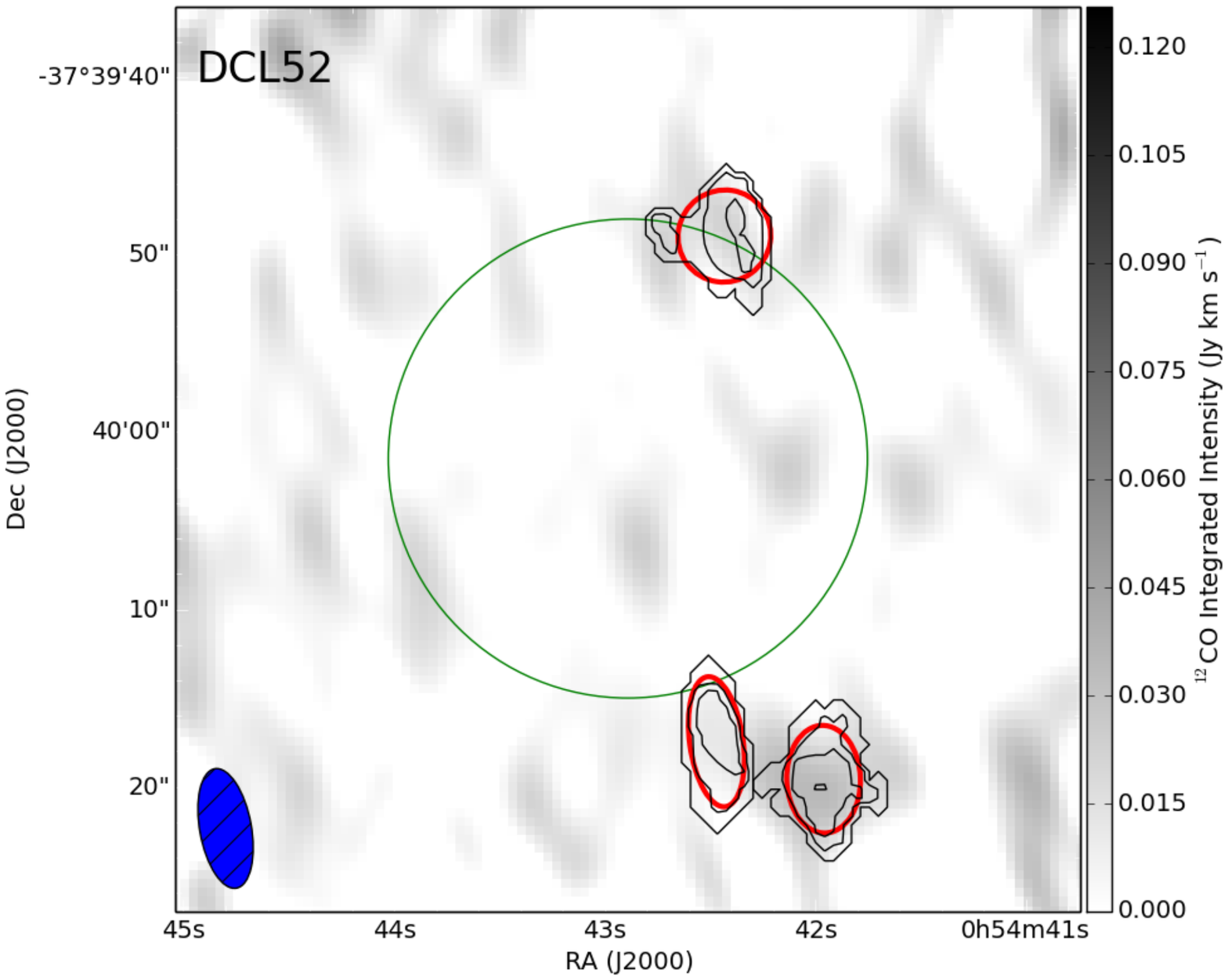}}
\subfigure{\includegraphics[trim={0 2.1in 0 2.1in},clip,width=0.33\linewidth]{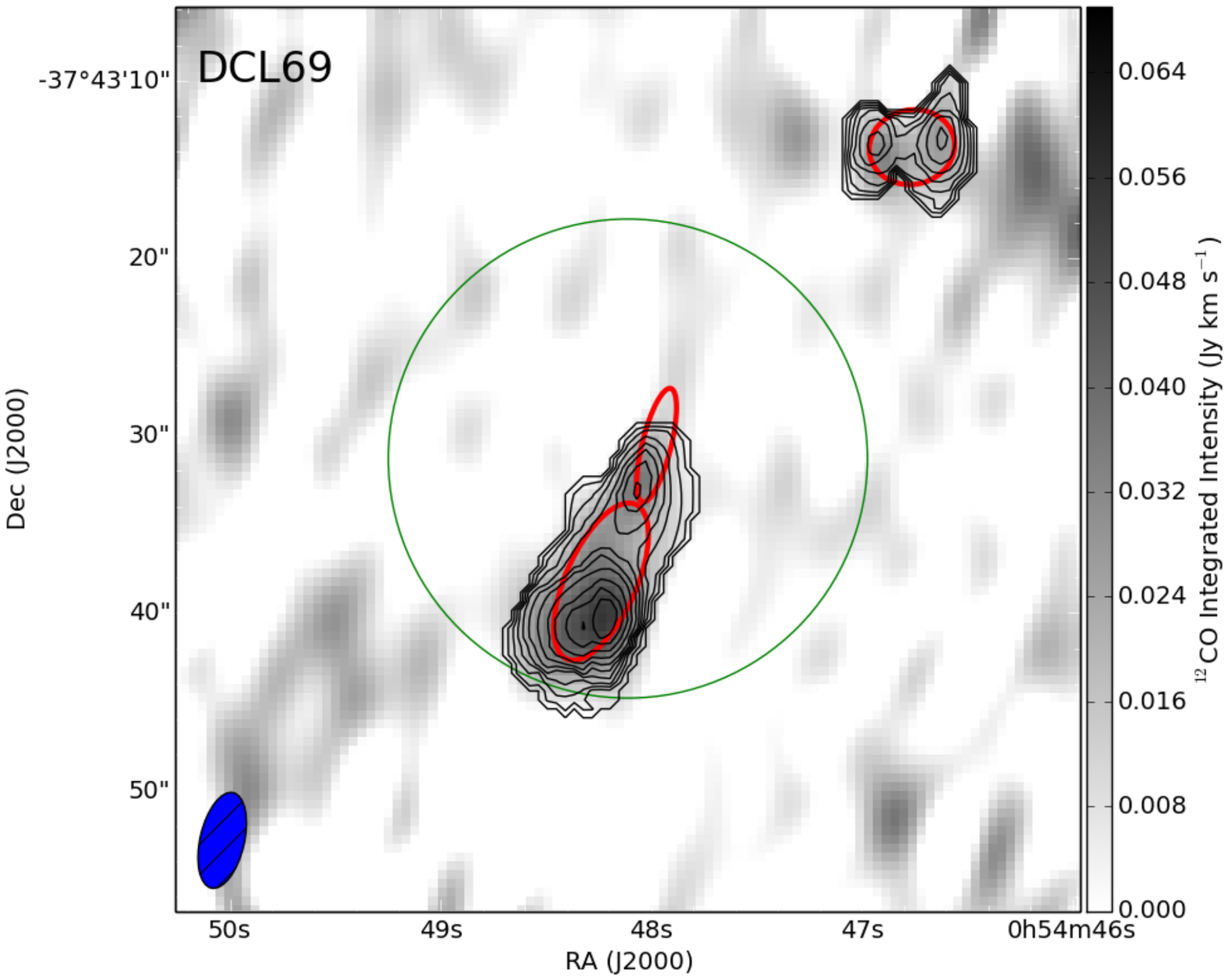}}
\subfigure{\includegraphics[trim={0 2.1in 0 2.1in},clip,width=0.33\linewidth]{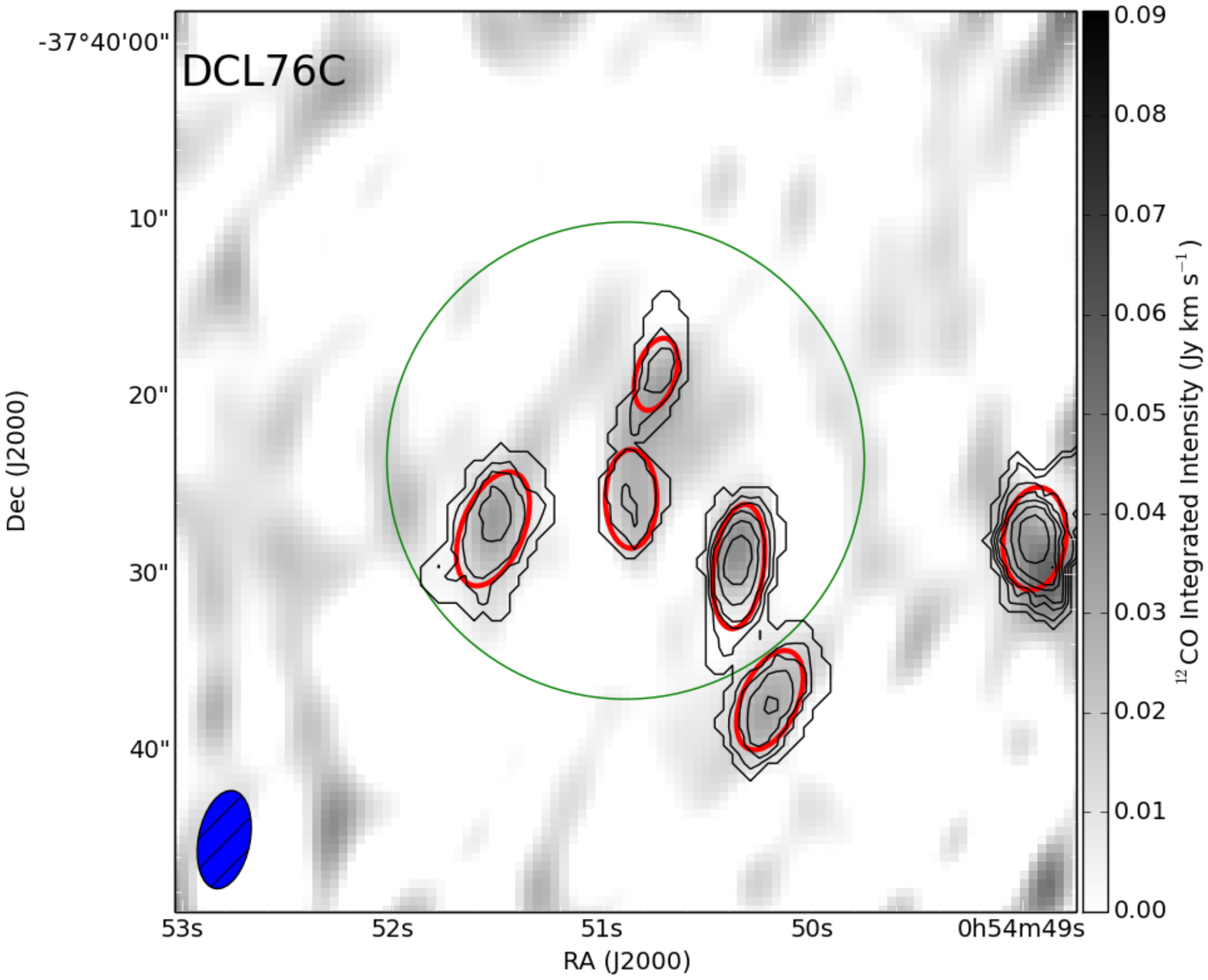}}
\subfigure{\includegraphics[trim={0 2.1in 0 2.1in},clip,width=0.33\linewidth]{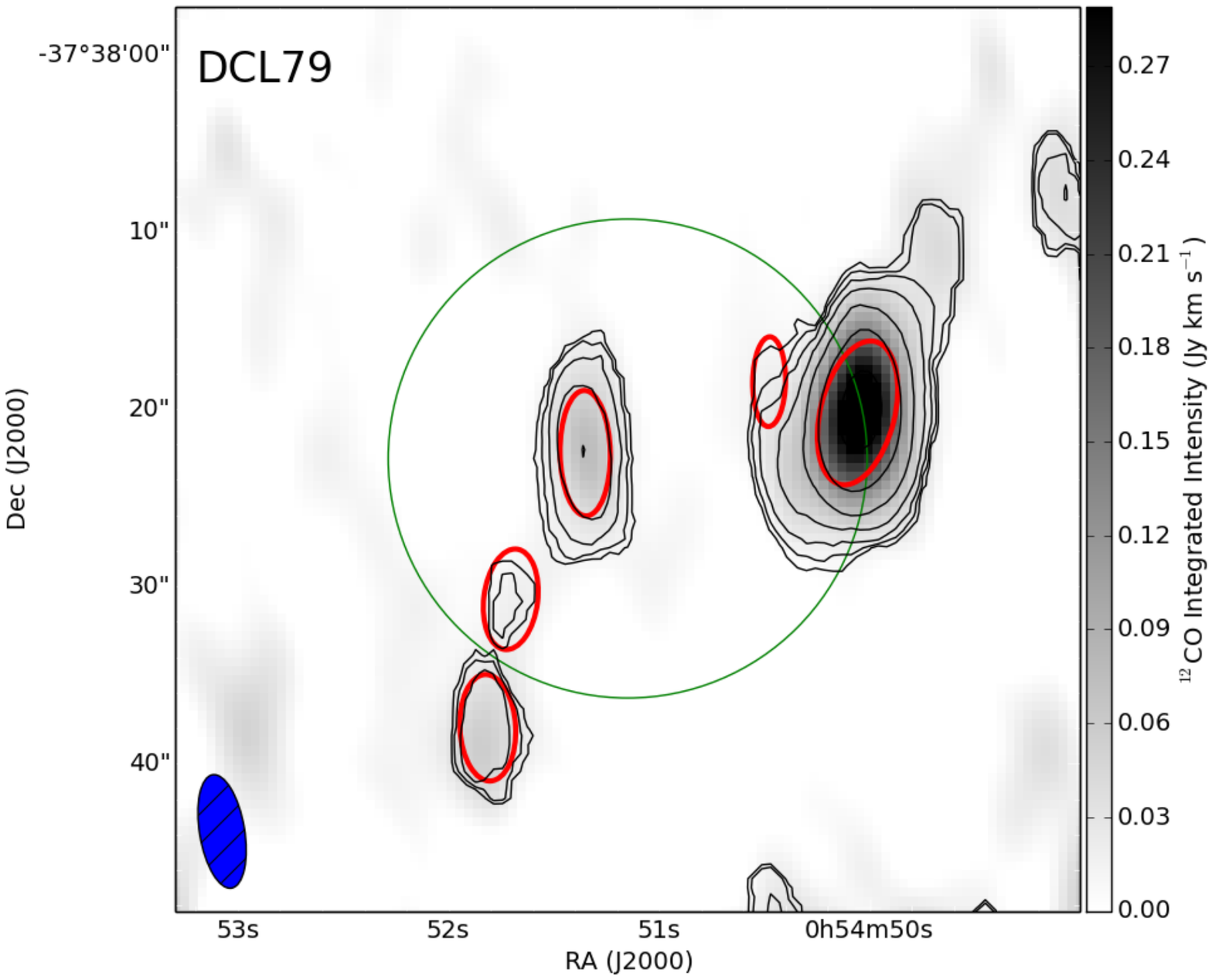}}
\subfigure{\includegraphics[trim={0 2.1in 0 2.1in},clip,width=0.33\linewidth]{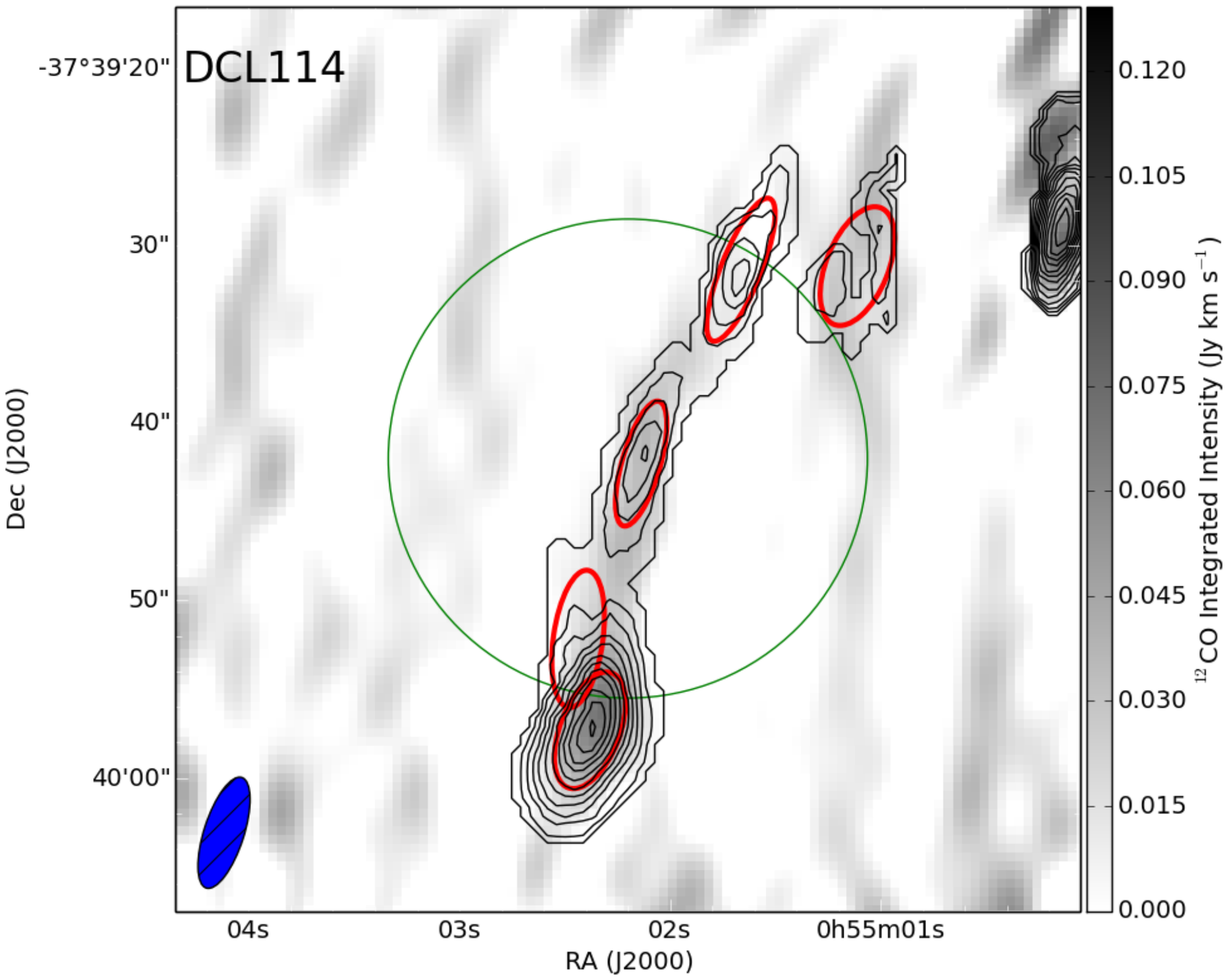}}
\subfigure{\includegraphics[trim={0 2.1in 0 2.1in},clip,width=0.33\linewidth]{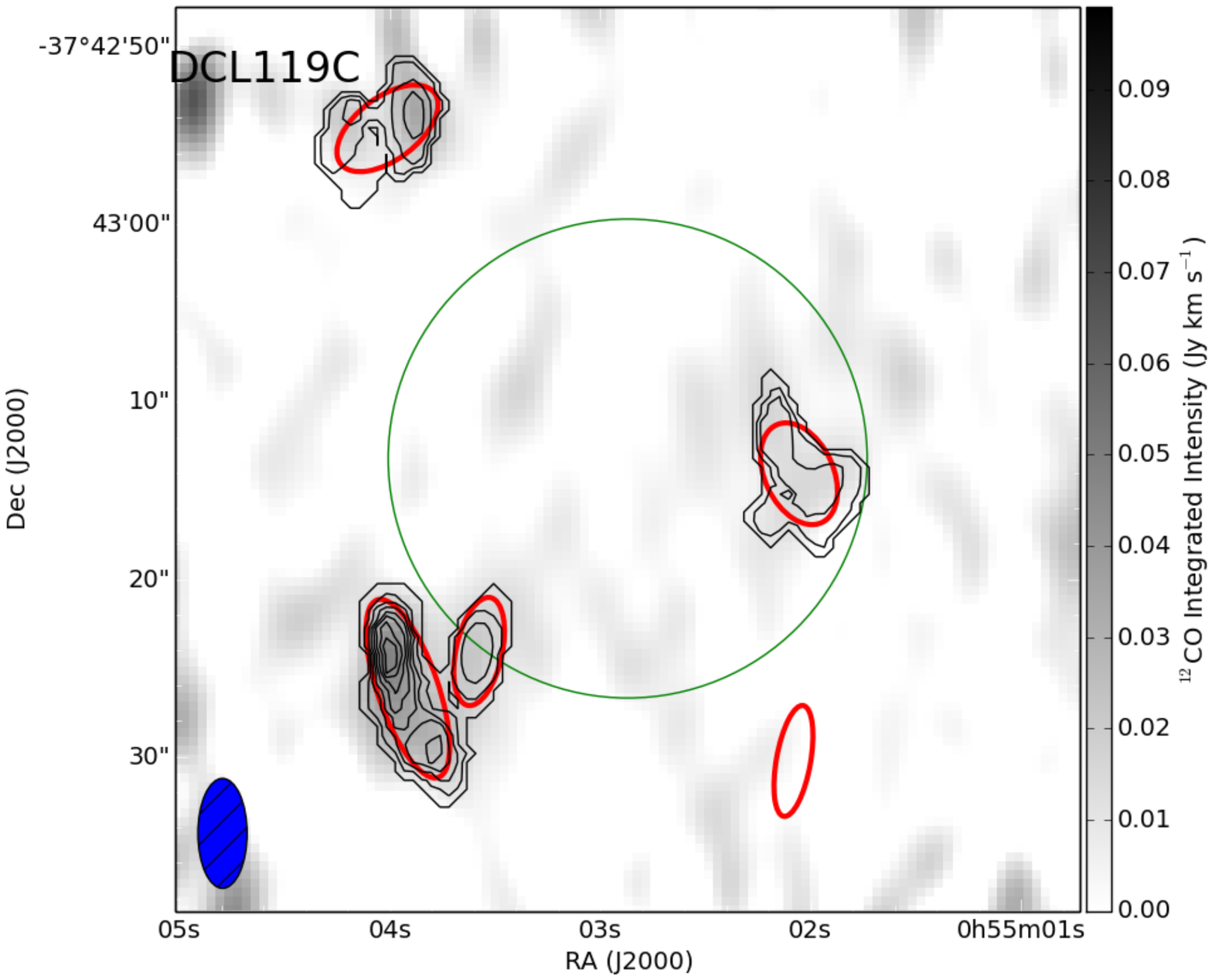}}
\subfigure{\includegraphics[trim={0 2.1in 0 2.1in},clip,width=0.33\linewidth]{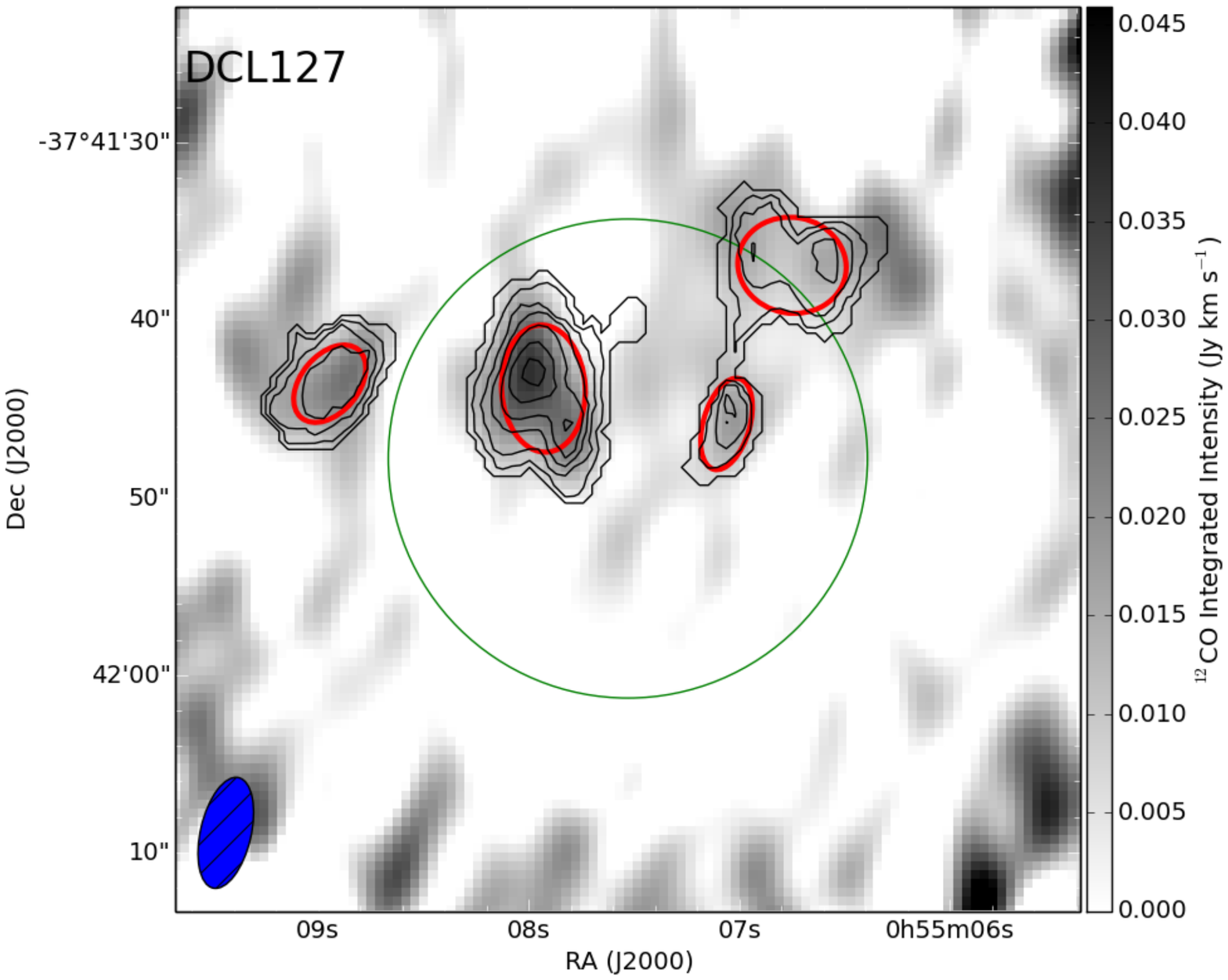}}
\captcont{{\CO} integrated intensity maps of the 9 regions observed with a single SMA pointing in grayscale, with \texttt{CPROPS}-identified GMCs overplotted as black contours. Contour levels are in integer multiples of the integrated intensity RMS noise beginning at $1\sigma$, except for DCL79 and DCL137, for which we plot contours spaced by powers of two times the RMS noise starting at $1\sigma$. The grayscale shows the integrated intensity in linear stretch from 0 Jy~{\kms}~to 80\% the image maximum. Red ellipses show the FWHM sizes and orientations of all \texttt{CPROPS} clouds. The synthesized beam in each image is indicated by the ellipse in the lower left. The green circle in each panel indicates the APEX $27\arcsec$ ($\sim250$~pc) FWHM pointing from \cite{Faesi:2014ib}.}
\label{fig:COmaps}
\end{figure*}

\begin{figure}[!htbp]
\centering
\includegraphics[trim={0 2.1in 0 2.1in},clip,width=\linewidth]{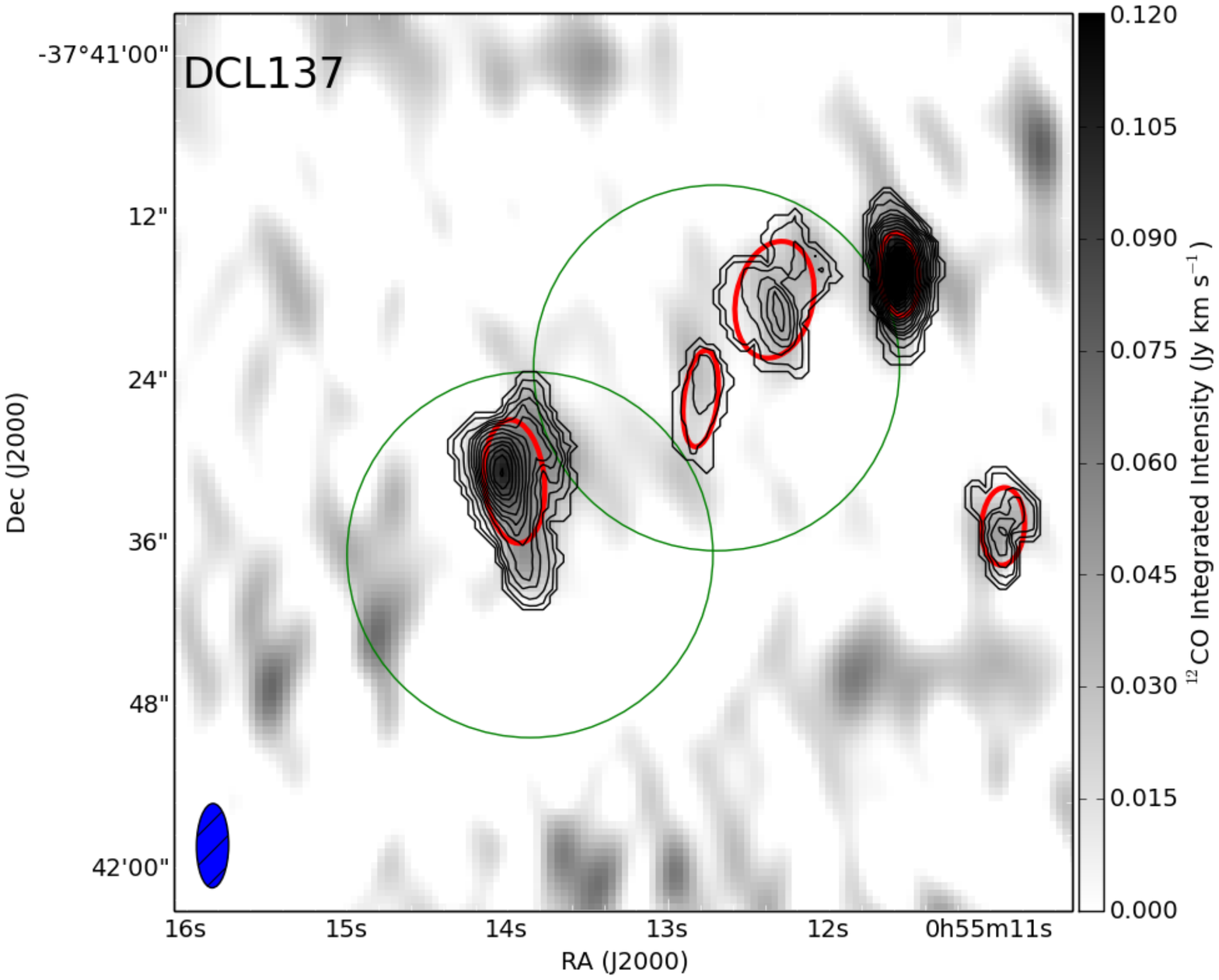}
\caption{(Continued.)}
\end{figure}

\subsection{GMC Physical Properties}

\begin{figure}[!tbp]
\includegraphics[trim={0 1.5in 0 1.5in},clip,width=\linewidth]{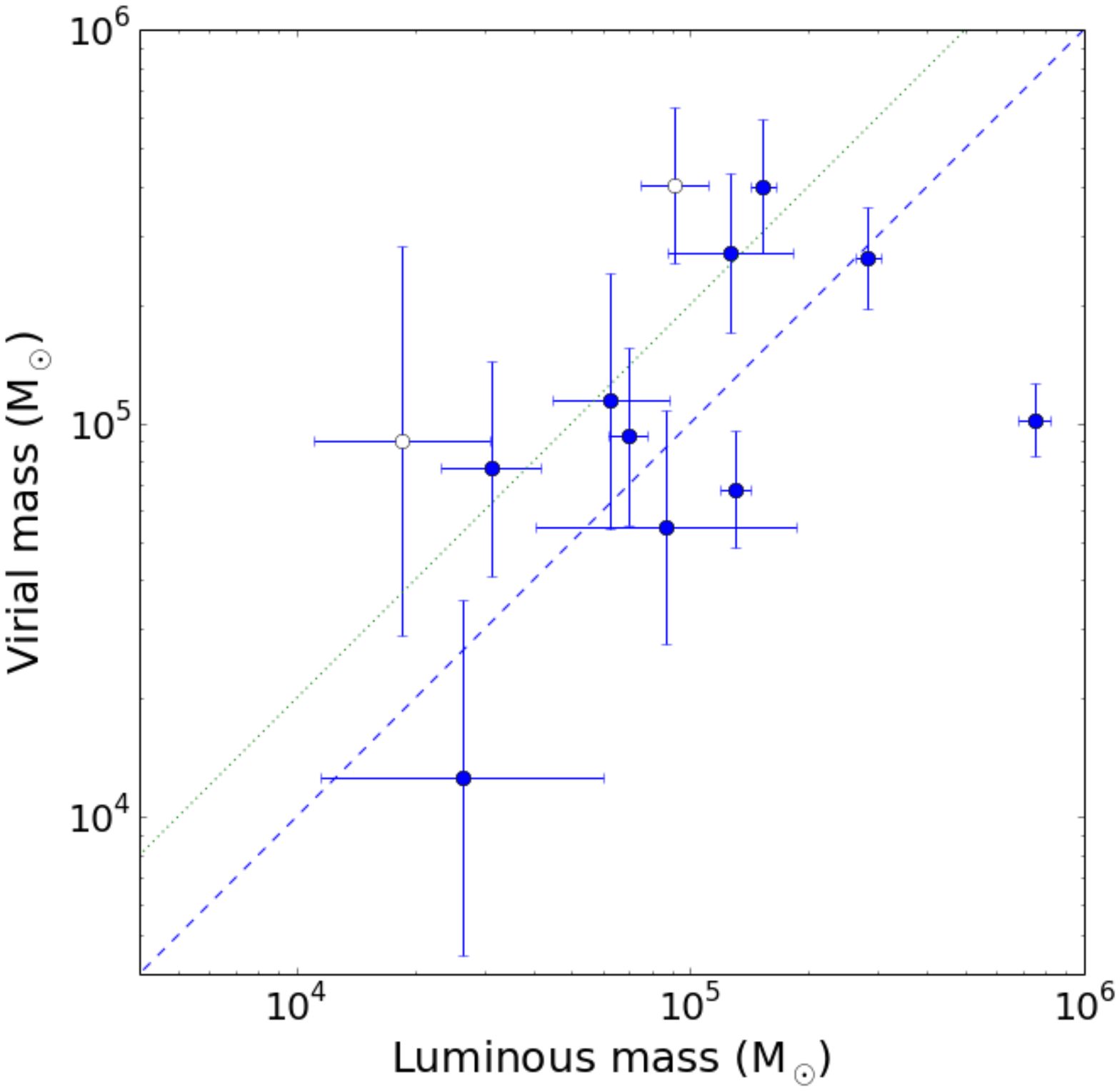}
\caption{GMC masses computed using CO luminosity and the X-factor ($M_{\rm lum}$) vs. virial masses $M_{\rm VT}$ for the resolved subsample. The dashed and dotted lines indicate a one-to-one and two-to-one relation, respectively, between the virial and luminous masses. Clouds with evidence for multiple components in their spectra are shown as open circles, while clouds that are well-fit by a single Gaussian are shown as solid circles. The majority (9/10) of single-component GMCs are bound or virialized, i.e. they have virial and luminous masses consistent with being within a factor of 2 of one another.}
\label{fig:masscomp}
\end{figure}

The GMCs in our sample have linewidths ranging from 1.8 to 8.3~{\kms}, (luminous) masses from several thousand to several hundred thousand {\Msun}, and deconvolved sizes of 9 to 42~pc (for those that are resolved) -- numbers that are within the range of GMC properties in other galaxies \citep[e.g.,][]{2008ApJ...686..948B}, as well as the Milky Way \citep{Solomon:1987uq}. The position angles of the resolved subsample of clouds after deconvolution range from $-75^{\circ}$ to $+90^{\circ}$, and do not appear to be correlated with the beam position angles. Note that due to symmetry, the full range of position angles is $-90^{\circ}$ to $90^{\circ}$, so the resolved subsample essentially spans this whole space. Figure~\ref{fig:masscomp} shows a general agreement between the luminous and virial masses of the resolved subsample, with the exception of clouds whose spectra show evidence for multiple unresolved velocity components (see below). For the majority of the GMCs, the ratio $M_{\rm VT}/M_{\rm lum}$ is consistent with being between one and two, suggesting that the clouds are approximately virialized. The clouds with virial masses more than a factor of two higher than their luminous masses appear a priori unbound, but may be confined by external pressure, but it is mostly likely that \texttt{CPROPS} has mistakenly identified multiple small clouds as a single larger object, and the linewidth of the falsely identified GMC is artificially increased due to the inflated velocity extent of the underlying unresolved clouds. The latter is certainly the case for the two clouds with obvious multiple-component spectra (DCL41-4 and DCL114-5; see Section~\ref{sec:spectra}). Note that DCL79-2 is likely an outlier primarily due to its size (and thus virial mass) being underestimated (see Section~\ref{sec:uncertainties}).

\subsection{Extracting GMC spectra}
\label{sec:spectra}

\begin{figure*}[tbp]
\includegraphics[trim={0.6in 2.1in 0.6in 2.2in},clip,width=\linewidth]{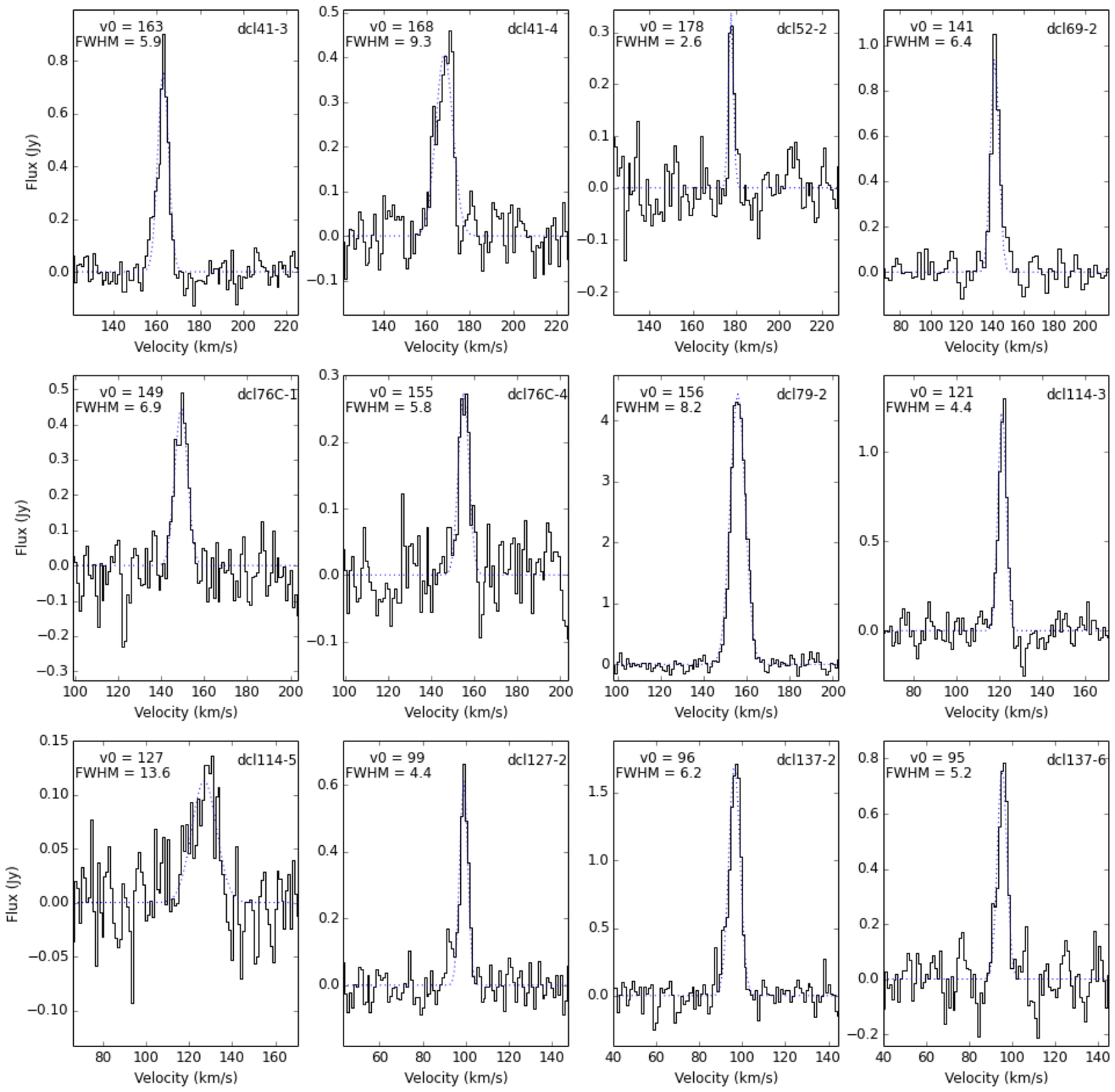}
\caption{Summed spectra (solid histograms) for the resolved subsample of GMCs along with Gaussian fits (dotted lines) to each spectrum. The region number and cloud number are indicated in the upper right of each panel, and the best-fit velocity centroid $v_0$ and velocity FWHM in the upper left. The majority of cloud spectra are relatively well-fit by a single Gaussian, with the notable exception of DCL41-4 and DCL114-5, which show clear evidence for a second velocity component at a comparable amplitude to the primary feature.}
\label{fig:spectra}
\end{figure*}

To further investigate whether the clouds extracted using \texttt{CPROPS} are individual GMCs (to the limit of our spectral and spatial resolution), we have extracted spectra summed over the spatial regions assigned to each cloud. Figure~\ref{fig:spectra} shows these spectra for the 12 GMCs in the resolved subsample. We conducted a single Gaussian nonlinear least-squares fit to each spectrum with the python \texttt{SCIPY.OPTIMIZE.CURVE\_FIT} routine, using the spectral flux maximum, the \texttt{CPROPS} cloud central velocity, and \texttt{CPROPS} velocity dispersion as initial guesses for the Gaussian amplitude, velocity centroid, and velocity width parameters, respectively. Ten of the twelve clouds are very well fit by a Gaussian function, while two GMCs, DCL41-4 and DCL114-5 are reasonably fit but also show evidence for a second velocity component that causes the single-function fit quality to suffer. The linewidths computed by \texttt{CPROPS} for these clouds are thus likely overestimates, as the voxels in the cloud contribute emission from multiple line-of-sight components that are inaccurately taken to be the distribution from which the velocity second moment is computed. Therefore the virial masses for these clouds are likely overestimated as well. These candidate multiple-component GMCs are indicated in all figures with open symbols to delineate them from the rest of the sample.

\subsection{CO Isotopologues}
\label{sec:isotopes}

\begin{figure*}[tbp]
$\begin{array}{cccc}
\includegraphics[trim={0 0.5in 0 2in},clip,width=0.25\linewidth]{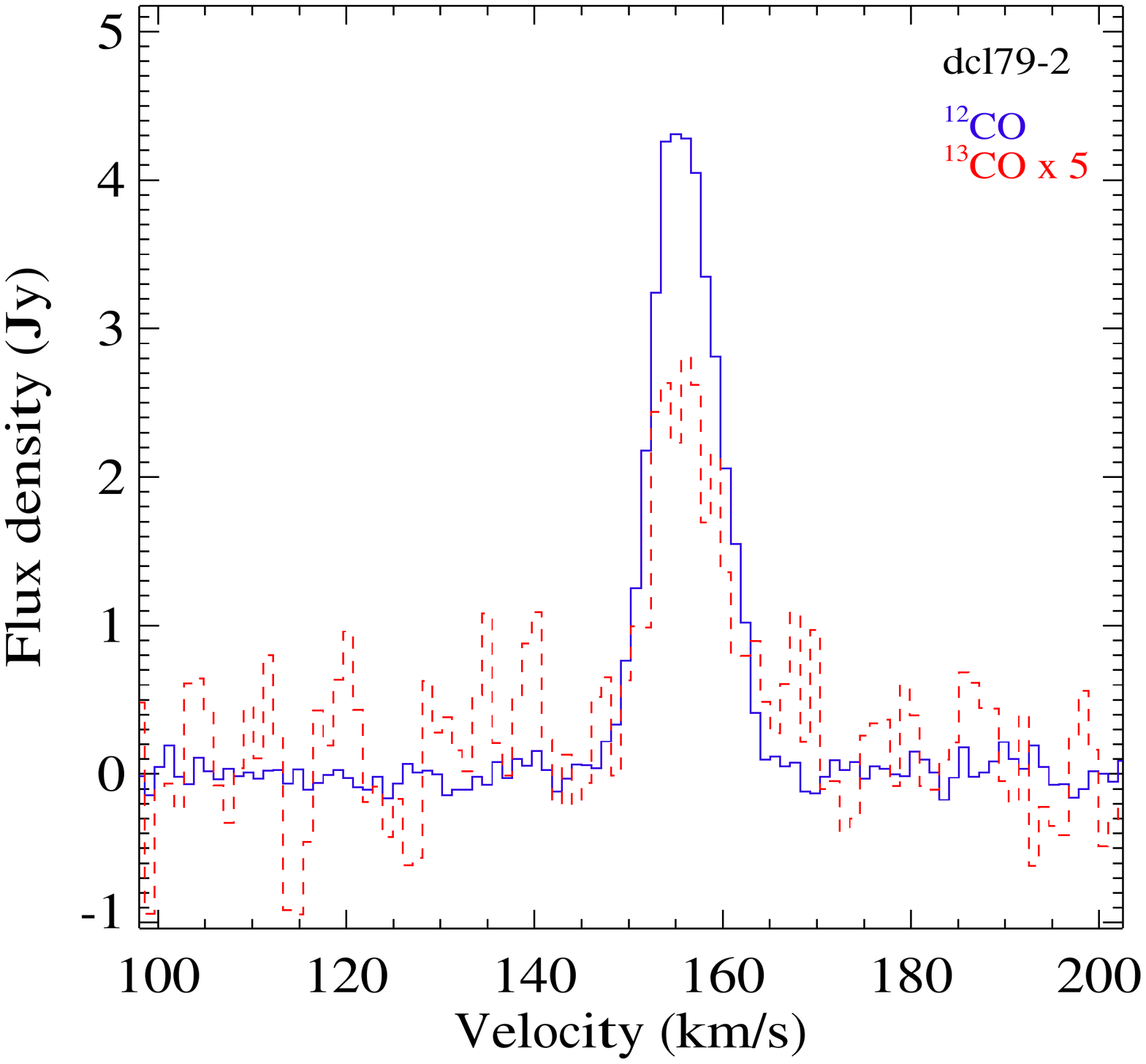} &
\includegraphics[trim={0 0.5in 0 2in},clip,width=0.25\linewidth]{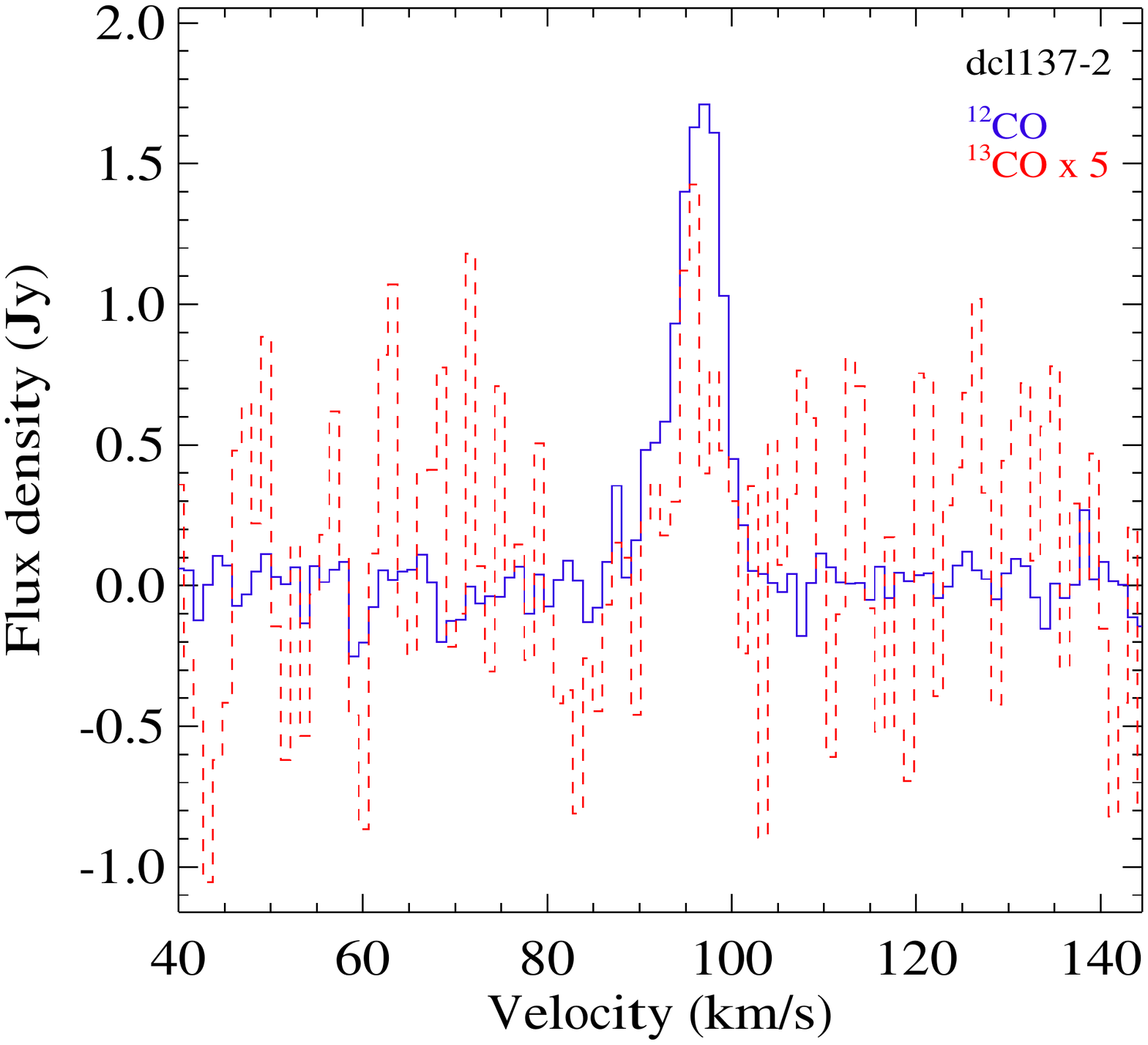} &
\includegraphics[trim={0 0.5in 0 2in},clip,width=0.25\linewidth]{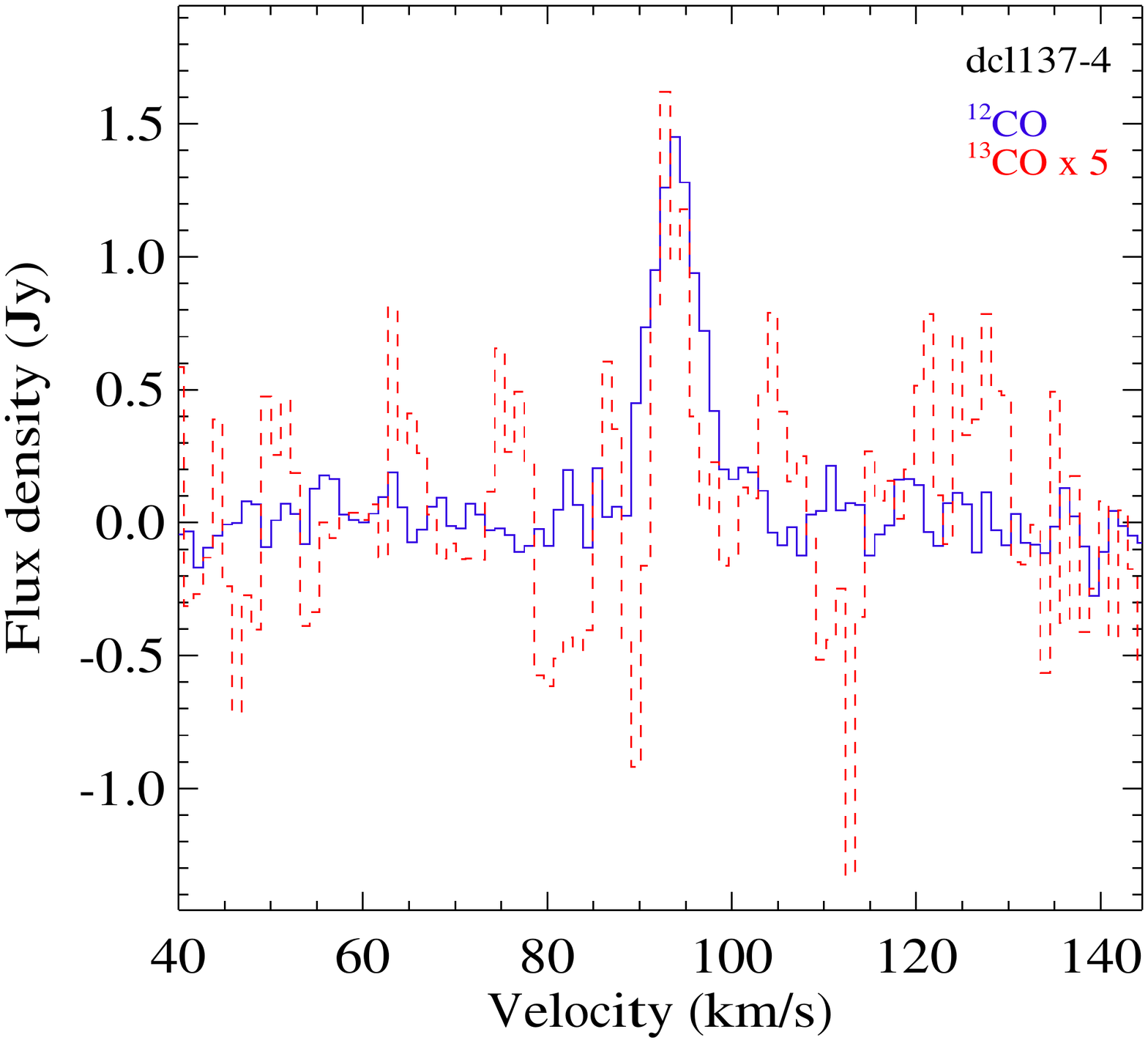} &
\includegraphics[trim={0 0.5in 0 2in},clip,width=0.25\linewidth]{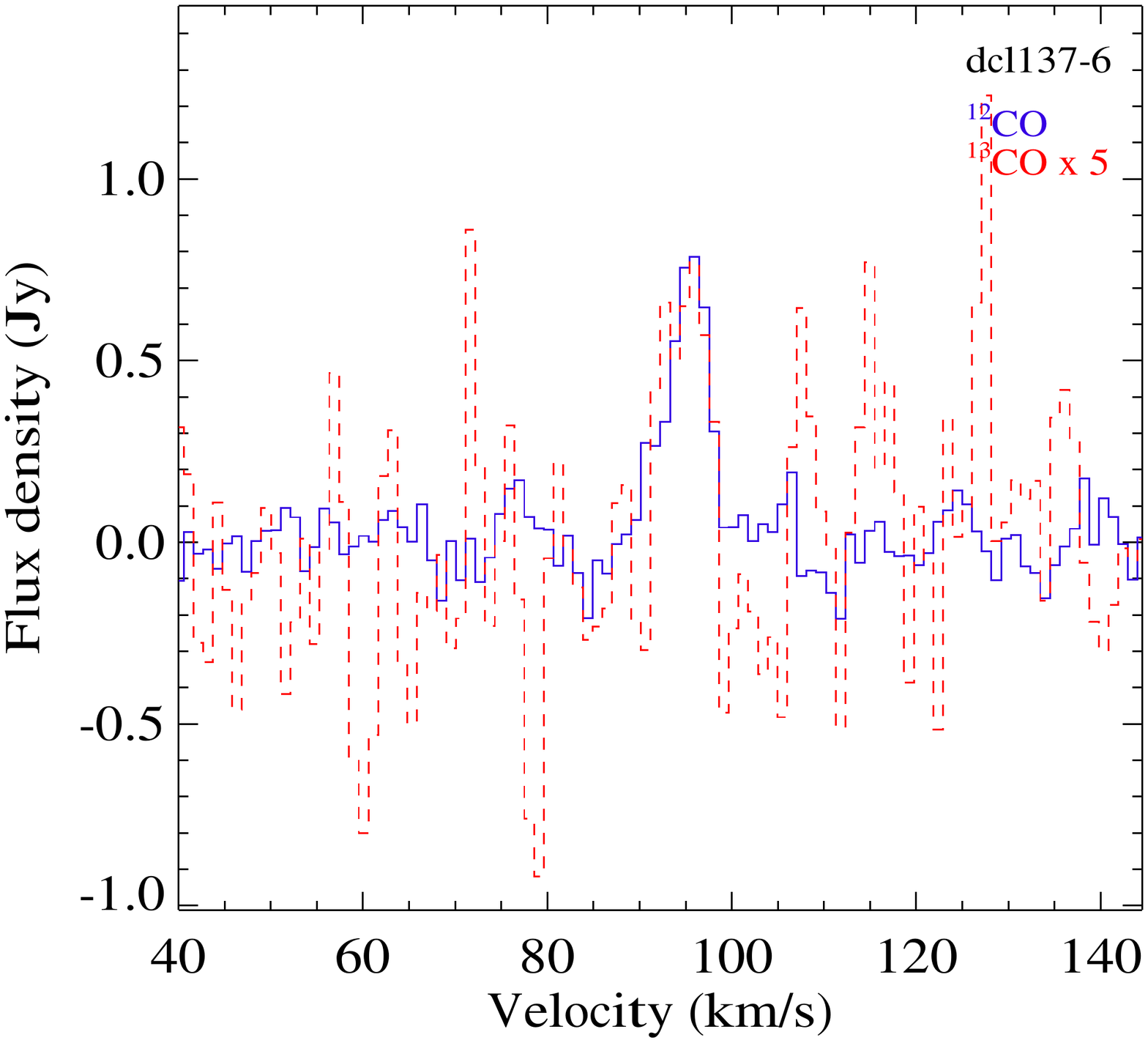} \\
\end{array}$
\caption{{\CO} (solid blue lines) and $^{13}$CO spectra (dashed red lines) of the four GMCs in which we significantly detect $^{13}$CO, DCL79-2, DCL137-2, DCL137-4, and DCL137-6. The $^{13}$CO spectra were multiplied by a factor of 5 to aid visibility. Spectra were constructed by integrating the emission over all pixels in the cloud as defined by {\CO}. The mean {\CO}/$^{13}$CO line ratio is 6.2.}
\label{fig:12and13co}
\end{figure*}

Our spectral setup included $^{13}$CO($J=2-1$) and C$^{18}$O($J=2-1$) for most of our observations (though notably not DCL88-69). We detect $^{13}$CO emission in only two regions, DCL79 and DCL137, and only near the brightest peaks in {\CO}. We do not detect C$^{18}$O at any significance in any region. Based on the RMS noise in the C$^{18}$O spectrum of DCL79-2 (our brightest cloud in {\CO}), we estimate a conservative lower limit on the {\CO}/C$^{18}$O line ratio of 19 (3$\sigma$). We present in Figure~\ref{fig:12and13co} the $^{13}$CO spectra of the four GMCs (DCL79-2, DCL137-2, DCL137-4, and DCL137-6) in which it was detected overplotted on their {\CO} spectra. We compute the {\CO}/$^{13}$CO line ratio, which we define as $F_{^{12}\rm{CO}}/F_{^{13} \rm{CO}}$, on a channel-by-channel basis. Over the range of velocities within twice the {\CO} FWHM where both {\CO} and $^{13}$CO emission are greater than two times their respective RMS values, the median {\CO}/$^{13}$CO line ratio ranges from 5.1 to 7.7 in these four GMCs. The mean {\CO}/$^{13}$CO line ratio across the sample of four clouds is 6.2 -- very similar to the ratio measured in Milky Way GMCs \citep[5.5;][]{1979ApJ...232L..89S}.

\section{Discussion}
\label{sec:discussion}

\subsection{GMC Mass Spectrum}

\begin{figure}[tbp]
\includegraphics[trim={0 1.5in 0 1.5in},clip,width=\linewidth]{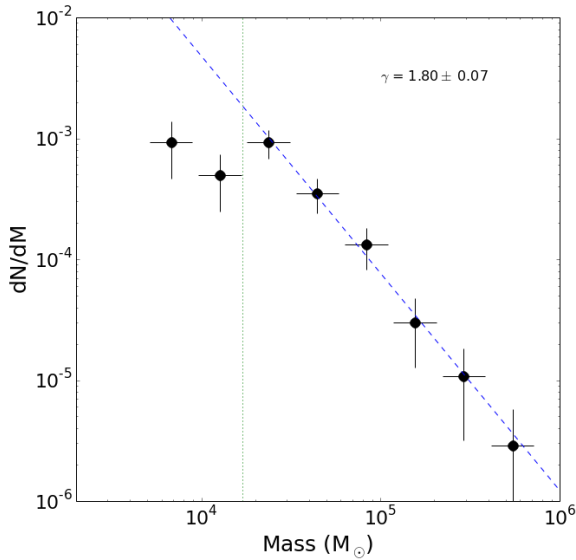}
\caption{GMC mass spectrum for the full sample of 45 clouds in our NGC 300 sample. GMC luminous masses were binned into logarithmically spaced bins of width 0.26 dex. A fit to the distribution using all bins fully above the completeness limit of $1.7 \times 10^4~\Msun$ (indicated by the vertical dotted line) yields a slope of $1.80\pm0.07$; varying the bin width over a reasonable range results in similar slopes, to within a factor of $\sim3$ times the quoted uncertainty.}
\label{fig:SMAmassspec}
\end{figure}

The distribution of GMCs by mass is a critical diagnostic of a GMC population and provides clues to GMC formation pathways. It may also be one of the only characteristics of GMC populations that varies between galaxies \citep[e.g.,][]{Rosolowsky:2005gt}. In this section we will discuss the GMC differential mass spectrum, which is expressed in the form $dN/dM \propto M^{-\gamma}$, in our sample of 45 clouds in NGC~300. The slope $\gamma$ is the critical parameter, and describes the relative fraction of clouds at low and high mass. For this analysis we choose to use the CO-derived masses $M_{\rm lum}$ over virial masses for several reasons: (1) we have a reliable estimate of $M_{\rm lum}$ for all 45 GMCs in our sample, unlike $M_{\rm VT}$ which is only available for the resolved subsample; (2) $M_{\rm lum}$ does not require assumptions about the dynamical state of the GMC; (3) for convention with existing resolved studies of GMCs in the Milky Way and other galaxies \citep[e.g.,][]{2001ApJ...551..852H,2003ApJS..149..343E}. We follow a similar procedure to \cite{Faesi:2014ib}, binning the sample logarithmically by mass in $\sim0.26$~dex bins. This corresponds to about twice the mean fractional uncertainty in $M_{\rm lum}$ across the sample. We estimate the value of each bin by dividing the number of clouds by the (linear) width of the bin. To calculate the slope of the mass spectrum, we fit the histogram above our estimated completeness limit using linear least squares fitting, with the uncertainties on each bin assumed to be Poissonian. We estimate the mass completeness limit by multiplying our $3-\sigma$ column density sensitivity limit of $12~\Msun$~pc$^{-2}$ by the average synthesized beam area of 1400~pc$^2$, arriving at a cloud mass sensitivity of $1.7 \times 10^4~\Msun$. We include all bins entirely above this completeness limit in this fit, and arrive at a slope of $\gamma=1.80\pm0.07$. We present the GMC mass spectrum in our eleven SMA-observed regions in Figure~\ref{fig:SMAmassspec}. Changing the bin size between 0.21 (the minimum bin size to have at least one GMC per bin) and 0.41 dex (about three times the average fractional uncertainty in $M_{\rm lum}$) results in derived slopes ranging from $\gamma = 1.64$ to 1.90, with uncertainties on $\gamma$ of about 0.2 and no systematic trend in slope with bin size.

The slope ($\gamma = 1.8$) we derive here is intermediate between that of the inner Milky Way \citep[$\gamma = 1.5$;][]{Solomon:1987uq} and outer Milky Way \cite[$\gamma = 2.1$;][]{2001ApJ...551..852H}. This suggests that in NGC~300, both high and low mass GMCs contribute significant amounts of molecular mass, and likely star formation, to the galaxy. We unfortunately do not have sufficient statistics to divide the sample by galactocentric radius and determine if there is a real difference in the mass spectrum slope between the inner and outer regions of the galaxy, as there appears to be in the Milky Way.

Comparing to other galaxies in which resolved GMC measurements are available, and assuming our sample of GMCs is representative of the full GMC population of NGC~300, we find that NGC~300's mass spectrum has a similar slope to that of the Large Magellanic Cloud \citep[$\gamma=1.75$;][]{2008ApJS..178...56F}, but a shallower slope than M33 \cite[$\gamma=2.6$;][]{2003ApJS..149..343E} when taken as a whole. However, note that in the inner regions of M33 (within 4 kpc, which is also the largest galactocentric radius in our sample in NGC~300), the slope is significantly shallower \citep[$1.6$ in the inner 2 kpc, and $\sim 2.1-2.3$ between 2 and 4 kpc;][]{2007ApJ...661..830R,Gratier:2012km}. Our derived slope is also shallower than that of the mass spectrum across M51 ($\gamma=2.3$), but again similar to the slope in M51's spiral arms \citep[$\gamma \approx 1.8$;][]{2014ApJ...784....3C}. Since we are specifically targeting regions with active star formation, our sample likely better resembles the inner galaxy and spiral arm samples in the galaxies discussed above rather than the more diffuse outer Milky Way or M33. As \cite{2014ApJ...784....3C} point out, galactic environment can play a strong role in shaping the GMC mass spectrum, to the point that the slope changes significantly with environment (for example, becoming steeper with increasing galactocentric radius). This can be interpreted theoretically as a change in the relative formation and destruction times of molecular clouds such that in more diffuse regions, the lack of availability of dense material leads to an increase in formation time while the destruction time remains relatively constant \citep{Inutsuka:2015gm}.

\cite{Faesi:2014ib} found the mass spectrum of GMC Complexes (at 250 pc scales) in NGC 300 to have a very steep slope of $\gamma=2.7 \pm 0.5$ above a completeness limit of $\sim1.5 \times 10^5~\Msun$. However, they did not resolve individual clouds, and we can see from our data that each APEX GMC Complex breaks up into two or more GMCs at high resolution. Furthermore, only 4 of the 45 clouds in our SMA sample have individual masses higher than the completeness limit from \cite{Faesi:2014ib}), so we are clearly probing different mass regimes as well as size scales. Further studies of NGC~300 will be necessary to connect these differing regimes and assess whether the GMC and GMC Complex mass spectra are related.

\subsection{The Larson Relations in NGC 300}

\cite{Larson:1981vv} first demonstrated the existence of empirical relations between the size, linewidth, and luminosity of GMCs in the Milky Way. In the first such relation, the linewidth scales with approximately the square root of the size \citep{Solomon:1987uq}, i.e.

\begin{equation}
\Delta V \approx 1.69 R^{0.5}.
\label{eqn:sizelinewidth}
\end{equation}

\begin{figure}[!tbp]
\includegraphics[trim={0 1.5in 0 1.5in},clip,width=\linewidth]{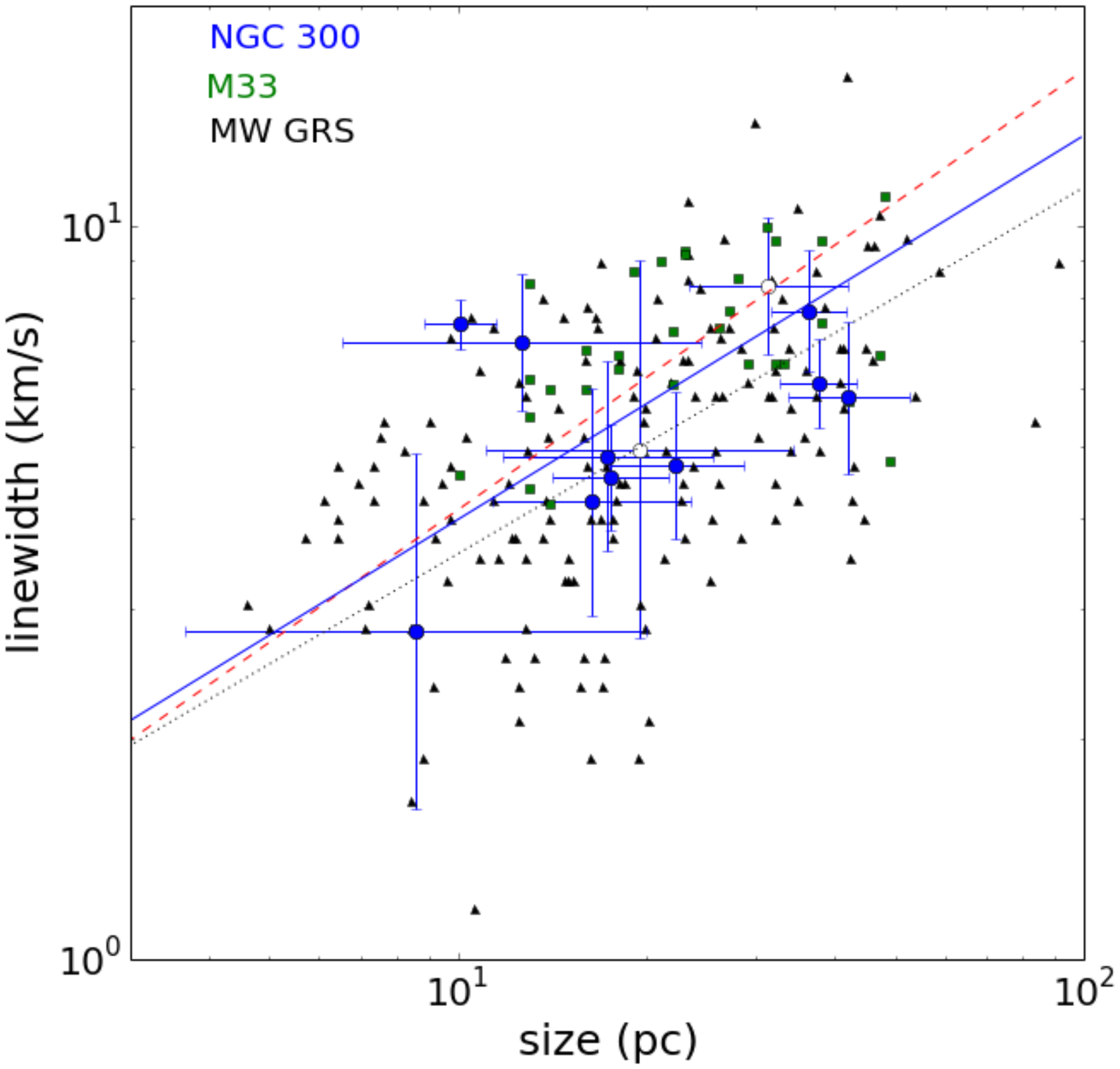}
\caption{Linewidth-size relation for the resolved subsample of GMCs, compared to Milky Way clouds from the GRS \citep[][black triangles]{2006ApJS..163..145J,Heyer:2009ii} and M33 clouds \citep[][green squares]{2003ApJ...599..258R}. The black dotted line indicates the $\Delta V \propto R^{0.5}$ relation, with the intercept from Equation~(\ref{eqn:sizelinewidth}) recalculated from the GRS data, while the red dashed line traces the extragalactic size-linewidth relation from \cite{2008ApJ...686..948B}. The blue solid line shows the best-fit relation from the NGC~300 data alone, $\Delta V \propto R^{0.52\pm0.20}$. The NGC~300 cloud properties are consistent with those in the Milky Way and M33.}
\label{fig:Larson1}
\end{figure}

\noindent This relation reflects the equilibrium conditions of supersonic turbulence in molecular clouds: turbulence appears to be injected on large scales and cascades down to smaller scales, decreasing in kinetic energy as it does so. The mechanisms that inject and maintain this turbulence are still debated, and may include feedback from newly formed stars, energy from supernova explosions, and/or galaxy-scale effects including spiral shocks and shear instabilities. Regardless of the origins, it has somewhat surprisingly been shown that a size-linewidth relation holds amongst GMCs in many nearby spiral galaxies \citep[e.g.][]{2003ApJ...599..258R,2008ApJ...686..948B,2013ApJ...772..107D}, although M51 at least appears to be an exception \citep{2014ApJ...784....3C}. These results suggest that there is some universality in the turbulent equilibria of the molecular ISM in many galaxies, independent of spiral galaxy mass and morphology. We present the size-linewidth relation for the resolved subsample in NGC~300 in Figure~\ref{fig:Larson1}, compared with Milky Way GMCs from the Five Colleges Radio Astronomy Observatory Galactic Ring Survey \citep[GRS;][]{2006ApJS..163..145J,Heyer:2009ii} and the M33 GMC sample of \cite{2003ApJ...599..258R}. Note that the GRS sample consists of the same clouds from \cite{Solomon:1987uq} but with properties recomputed based on higher resolution data, and also utilizing $^{13}$CO observations instead of {\CO}. We take the GRS cloud properties to be accurate, and thus for the relation shown in Figure~\ref{fig:Larson1}, we have recomputed the intercept in Equation~(\ref{eqn:sizelinewidth}) using the GRS data. The NGC 300 GMCs appear consistent in their sizes and linewidths to those in the Milky Way and M33.

Taken alone, our data suggest an increasing trend between size and linewidth, and we perform an orthogonal distance regression on the logarithms of these quantities in the resolved subsample to determine the best-fit power law relation between them. To perform the fitting, we utilize the \texttt{SCIPY.ODR} package, which accounts for errors in both variables. We use the \cite{Solomon:1987uq} relation to provide initial guesses for the parameters $a$ and $b$ to the equation $\log{\Delta V} = a + b \log{R}$. The formal best fit relation in the NGC~300 clouds is
\begin{eqnarray}
\log{\Delta V} &=& (0.08 \pm 0.24) + (0.52 \pm 0.20) \log{R}, \rm{ i.e.,} \nonumber \\
\Delta V &=& 1.1 \, R^{0.52\pm0.20.}
\end{eqnarray}
This exponent, while uncertain, is fully consistent with the value of 0.5 found by \cite{Solomon:1987uq} for the Milky Way, and within the uncertainties of the \cite{2008ApJ...686..948B} value for nearby galaxies. Thus quantitatively as well as qualitatively, the size-linewidth relation in NGC~300 is in agreement with the general trend in disk galaxies in the local Universe, including the Milky Way. We do note that the relatively large uncertainties in \texttt{CPROPS}-derived size and linewidth and limited dynamic range in our sample make it difficult to draw firm conclusions as to the physical implications of the size-linewidth correlation beyond the general empirical trend discussed above. Further high-resolution studies in NGC~300 are needed to better constrain these GMC parameters and improve the dynamic range in cloud properties.

\begin{figure}[!tbp]
\includegraphics[trim={0 1.5in 0 1.5in},clip,width=\linewidth]{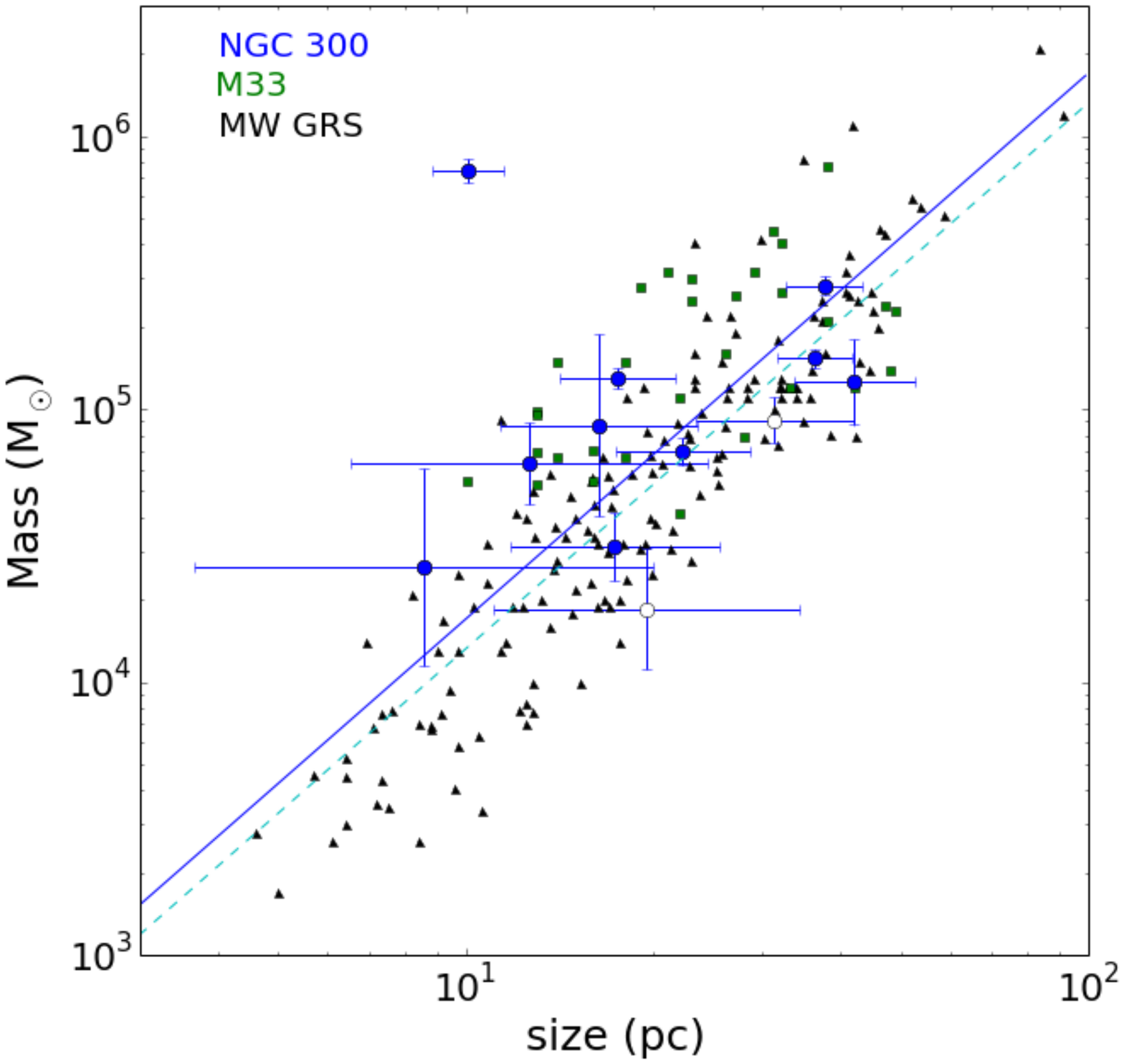}
\caption{Mass-size relation for the resolved subsample of GMCs, compared to Milky Way clouds from the GRS \citep[][black triangles]{2006ApJS..163..145J} and M33 clouds \citep[][green squares]{2003ApJ...599..258R}. The blue solid line represents the relation $M \propto R^2$ with the normalization set by the average surface density in our sample, $54~\Msun$~pc$^{-2}$. The cyan dashed line is the same relation with the average surface density from the Milky Way sample of \cite{2010A&A...519L...7L}, $41~\Msun$~pc$^{-2}$. GMC mass appears to increase systematically with size in the resolved subsample in NGC~300. The NGC~300 points also lie within the locus defined by the Galactic and M33 clouds, with the exception of DCL79-2, which may have an erroneously small computed size (see the text).}
\label{fig:Larson2}
\end{figure}

Larson's third relation (and the second we discuss here), which originally related the average volume densities and sizes of clouds, has received much attention recently in the literature \citep[e.g.,][]{Heyer:2009ii,2010A&A...519L...7L,2013MNRAS.436.3247K}. Larson found that the volume density $n_{\rm H2}$ scaled with size roughly as

\begin{equation}
n_{\rm H2} \propto R^{-1.1},
\end{equation}

\noindent i.e. the mass scales as size like $M \propto R^{1.9}$. One key implication is that the column density $\Sigma$ of GMCs, which goes as $M/R^2$, is approximately constant. \cite{2010A&A...519L...7L} showed that, using high-fidelity near-infrared extinction maps to trace cloud column densities, the value of this constant surface density is $\Sigma \approx 41~\Msun$~pc$^{-2}$, with very little scatter in the relation. This result is based on analysis in which the cloud is considered to extend out to an extinction threshold of $A_0 = 0.1$~mag in the $K$-band. \cite{Heyer:2009ii} find a similar value of 42~{\Msun}~pc$^{-2}$ when re-analyzing the GMCs from the \cite{Solomon:1987uq} sample in a consistent manner using GRS $^{13}$CO data. We present the mass-size relation in the resolved subsample in NGC~300 in Figure~\ref{fig:Larson2}, again compared to Milky Way GRS data and M33. We again find that the NGC~300 data are consistent with the mass-size relation in the Milky Way and M33. The median surface density in our sample is 54~$\Msun$~pc$^{-2}$, slightly higher than the Milky Way average~\citep{2010A&A...519L...7L}. We also note that surface densities in our sample range widely, from 15 to 125~$\Msun$~pc$^{-2}$, but with very large uncertainties (factors of one to a few). This discrepancy may reflect more on the large potential systematic uncertainties in deconvolved sizes (see Section~\ref{sec:uncertainties}), which enter the surface density calculation to the power of two, than on any intrinsic variation of surface density in GMCs. We note that one GMC, DCL79-2, appears to have an exceptionally high surface density, but this is likely a result of its size being underestimated (see Section~\ref{sec:uncertainties}).

\begin{figure}[!tbp]
\includegraphics[trim={0 1.5in 0.7in 1.5in},clip,width=\linewidth]{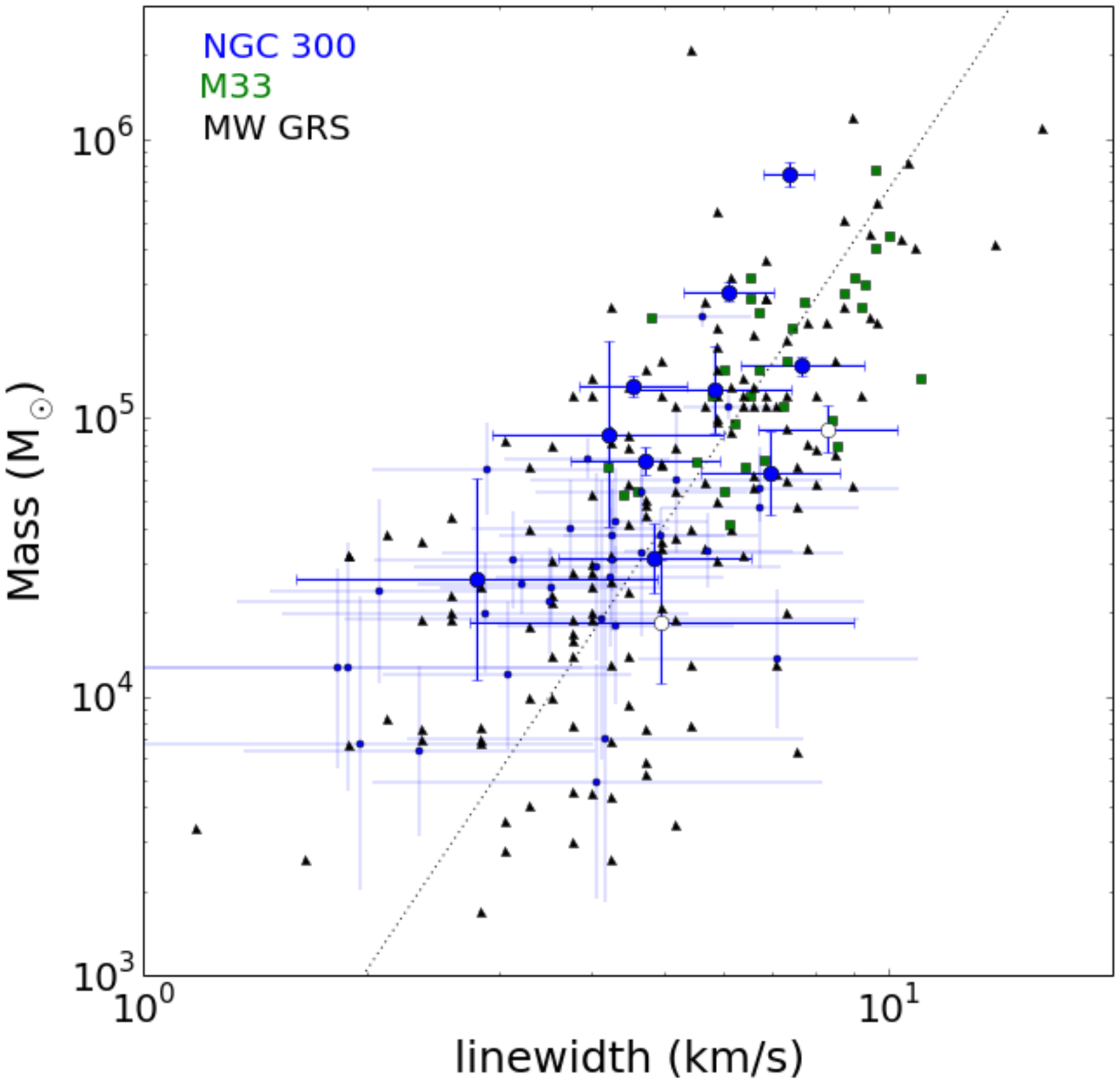}
\caption{Linewidth-mass relation for the resolved subsample of GMCs, compared to Milky Way clouds from the GRS \citep[][black triangles]{2006ApJS..163..145J} and M33 clouds \citep[][green squares]{2003ApJ...599..258R}. The large blue points represent the resolved subsample of NGC~300 clouds, while the small blue points are the unresolved NGC~300 clouds. The black dotted line traces the relation $M = 65.6 \, (\Delta V)^4$ from \cite{Solomon:1987uq}. Similar to the case for the other two relations, the linewidth-mass relation in the NGC~300 resolved subsample appears consistent with the relation observed in the Milky Way and M33.}
\label{fig:Larson3}
\end{figure}

We present the final Larson's relation, the mass-linewidth relation, for NGC~300 in Figure~\ref{fig:Larson3}, As only two of the three Larson relations are independent, it follows naturally that since the NGC~300 resolved subsample shows similar behavior to GMCs in the size-linewidth and mass-size relations in other galaxies (including the Milky Way), it also is similar in the mass-linewidth relation. We attempted to fit power law functions to the mass-size and mass-linewidth relations, but the large uncertainties in these quantities led to indeterminate fits.

\subsection{Comparing Interferometric and Single Dish Measurements}

\begin{figure*}[!tbp]
\centering
\includegraphics[trim={0.2in 1.4in 0.2in 1.4in},clip,width=0.8\linewidth]{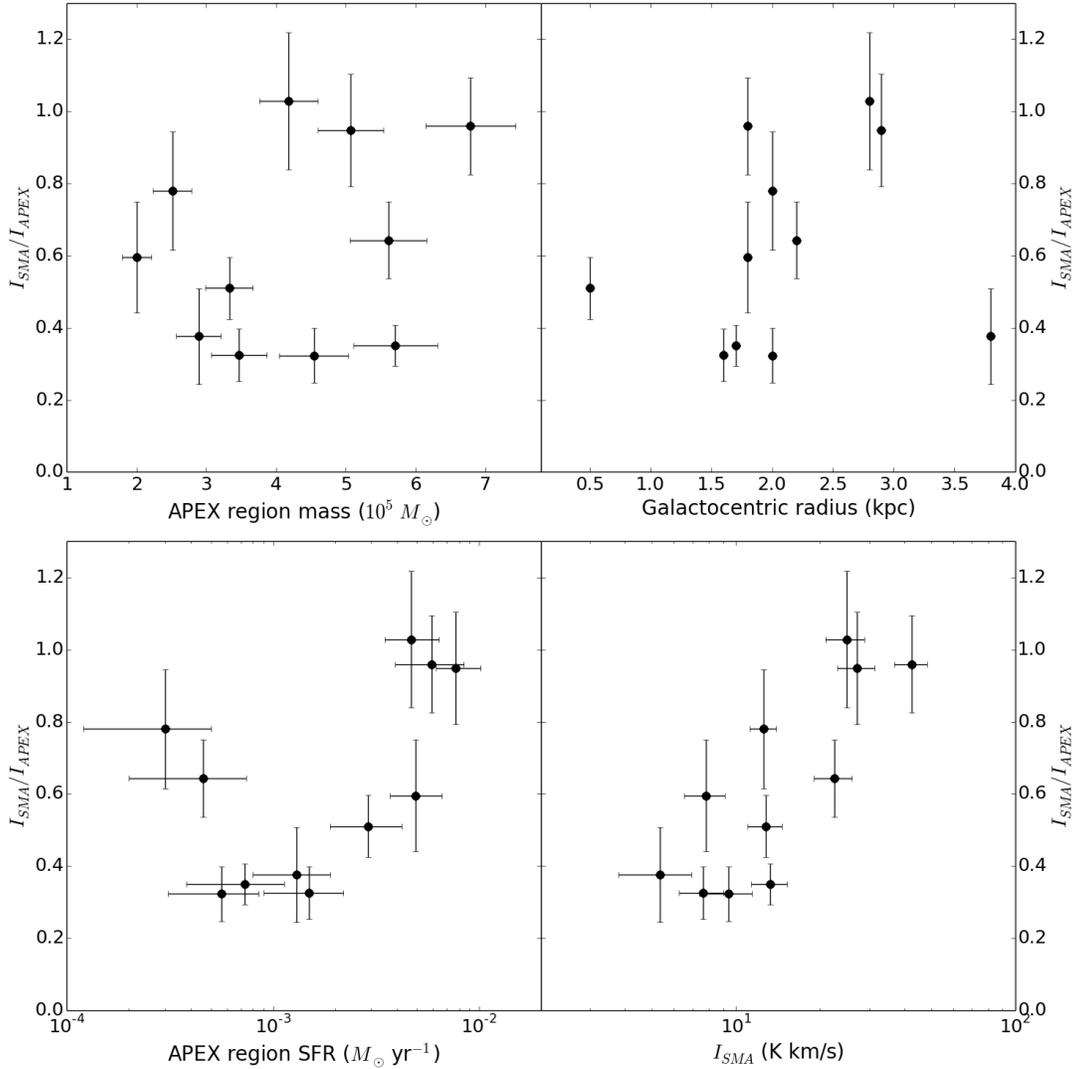}
\caption{$I_{\rm SMA}/I_{\rm APEX}$, the fraction of the APEX CO integrated intensity recovered by the SMA as a function of APEX region molecular mass (upper left), galactocentric radius (upper right), 250 pc-scale star formation rate (SFR; lower left), and total molecular mass in GMCs (lower right) from this study. There is no clear correlation between $I_{\rm SMA}/I_{\rm APEX}$ and APEX molecular mass or galactocentric radius, but the fraction appears to correlate with $I_{\rm SMA}$. A weak trend is also present with the SFR for SFR $\gtrsim 5 \times 10^{-4}~\Msun$~yr$^{-1}$.}
\label{fig:SMAAPEX}
\end{figure*}

With both pure interferometric and single dish data of the same regions, we can compare our results directly with those of the APEX survey of \cite{Faesi:2014ib} to assess what fraction of the single dish flux we recover with the SMA. To make this comparison, we perform the following procedure on the SMA primary beam-corrected data cubes. First, we mask all pixels (each of which is a one-dimensional spectrum) in the data cubes that do not show significant CO emission. Specifically, we require for inclusion that a pixel have at least one velocity channel in the spectral region including the line exhibiting emission greater than twice the spectral RMS. The spectral region here is taken to include all channels within twice the APEX spectral FWHM on either side of the APEX line center, as determined by single Gaussian fits to the APEX spectra. The spectral RMS is calculated from the portion of the spectrum outside this region. This criterion ensures that we do not include pixels in the spectral sum that are purely noise. Next, we multiply each plane of the SMA primary beam-corrected data cube by the APEX beam profile, which we approximate as a $27\arcsec$ FWHM circular Gaussian, centered at the APEX pointing position. Then, we sum all unmasked pixels in each plane over a circular region the size of the SMA primary beam, and the sequence of all such planes forms the SMA spectrum. Finally, we use the trapezoidal rule to numerically integrate this spectrum over a velocity range encompassing twice the APEX spectral FWHM and centered at the APEX central velocity to determine the APEX-attenuated SMA integrated intensity, $I_{\rm SMA}$. The uncertainty $\sigma_{I_{\rm SMA}}$ on $I_{\rm SMA}$ is given formally \citep[e.g.,][]{Faesi:2014ib} by
\begin{equation}
\begin{split}
\sigma_{I_{\rm SMA}} &= \sigma_{F_{\rm SMA}}\sqrt{\Delta v \, w}  \\
				  &= \sigma_{F_{\rm SMA}} \Delta v \sqrt{N},
\end{split}
\end{equation}
where $N = w / \Delta v$ is the number of channels across $w$, the velocity range over which the integration is performed, and $\Delta v$ is the velocity width of a channel. $\sigma_{F_{\rm SMA}}$ is calculated as the quadrature sum of the line-free spectral RMS and the 20\% flux calibration uncertainty.

The fraction of the APEX integrated intensity recovered by the SMA, $\mathcal{F}_{\rm rec}$, is then finally $\mathcal{F}_{\rm rec} = I_{\rm SMA} / I_{\rm APEX}$ and the uncertainty is computed by propagating uncertainties in the standard way. We take $I_{\rm APEX}$ and the corresponding uncertainty directly from \cite{Faesi:2014ib}. The mean recovered fraction is $\langle \mathcal{F}_{\rm rec} \rangle = 0.62 \pm 0.12$, which demonstrates that our SMA observations do not generally recover the full emission detected by APEX. One likely cause could be the relative sensitivities of the APEX and SMA observations combined with our masking procedure. The APEX observations have a median RMS noise of 11~mK per 1.389~{\kms} channel \citep{Faesi:2014ib}, while these SMA observations have an RMS of 27~mK when scaled to the same channel velocity width. Thus there may be real CO emission (e.g., from low-mass GMCs) that contributes to the APEX spectrum that we are simply not sensitive enough to detect with the SMA. Observations with increased sensitivity and better angular resolution would be likely to detect such clouds if they exist (e.g., Rosolowsky et al. 2013). Another possibility is that there is diffuse and extended CO emission that is resolved out due to incomplete $u$,$v$ coverage of the SMA observations.  The shortest baseline of 16.4 m corresponds to a maximum recoverable size scale of 16~{\arcsec}, which is about 150 pc at the distance of NGC 300. Although we should recover most of the APEX emission with a characteristic scale size smaller than this, diffuse emission with $>150$~pc scales would be filtered out in our SMA observations. However, it seems implausible that diffuse emission could explain the entire discrepancy for the lowest $\mathcal{F}_{\rm rec}$ regions, as this would imply that more than 50\% of the CO emission detected by APEX is extended, diffuse emission. Indeed in the Milky Way, only about 25\% of the global molecular gas reservoir is in diffuse form~\citep{2016arXiv160100937R}, and much of this is likely less extended than 150~pc.

We note that $\mathcal{F}_{\rm rec}$ does vary on a region-by-region basis, from 0.32 (DCL88-52) to about unity (DCL88-137B) across the sample. We examine potential causes of this variation in $\mathcal{F}_{\rm rec}$ visually in Figure~\ref{fig:SMAAPEX}, which shows $\mathcal{F}_{\rm rec}$ as a function of four parameters: (1) the 250~pc scale APEX region molecular mass; (2) galactocentric radius; (3) the 250~pc scale SFR; and, (4) $I_{\rm SMA}$. $\mathcal{F}_{\rm rec}$ is flat with APEX region mass (which is proportional to $I_{\rm APEX}$). This indicates that there is no systematic trend in how much emission the SMA recovers with single dish integrated intensity. There is also no correlation in $\mathcal{F}_{\rm rec}$ with galactocentric radius. There is a hint of a trend is in the $\mathcal{F}_{\rm rec}$-SFR relation at moderate to high SFRs: above $5 \times 10^{-4}~\Msun$~yr$^{-1}$, the fraction of single dish emission recovered appears to increase with the region's SFR. Finally, there is a clear trend in $\mathcal{F}_{\rm rec}$ with $I_{\rm SMA}$. This latter trend may result from a systematic variation in the relative amounts of emission from large GMCs detected by both APEX and SMA and some combination of low mass GMCs and potentially some diffuse molecular gas detected only in the APEX observations. In regions with very bright CO emission (massive GMCs), emission from these bright regions dominates and so the relative contribution of the low-level emission is small, leading to a higher $\mathcal{F}_{\rm rec}$. Conversely, in regions with only moderate mass GMCs, the relative contribution of the undetected emission can be significant, as suggested by $\mathcal{F}_{\rm rec}$ lower than 50\%.

We also briefly speculate here on a potential physical origin for the weak $\mathcal{F}_{\rm rec}$-SFR trend at moderate to high SFRs. One explanation is that regions having the majority of their molecular gas in large, massive GMCs are more actively star-forming than regions where the majority of molecular gas is exterior to large GMCs. Feedback from active star formation is energetically capable of efficiently disrupting diffuse molecular gas and small, low mass GMCs in short timescales through a high flux of ultraviolet radiation. Thus in the most actively star-forming regions (those with the highest SFRs), feedback is effectively disrupting the majority of the diffuse gas or low-mass GMCs that were present, while the molecular gas in massive GMCs (i.e., those we have the sensitivity to detect) is shielded enough to remain in the harsh radiation field for significant timescales. In quiescent (low SFR) regions, diffuse gas and low-mass GMCs may be able to better survive due to the weaker radiation fields present. The disagreement with the trend for the very lowest SFRs (below $5 \times 10^{-4}~\Msun$~yr$^{-1}$) may be (1) a result of these regions being in an earlier (pre-star forming) state of evolution (and thus SFRs underestimated using the method that relies on tracers of young massive stars), but with the majority of their gas bound in GMCs that will soon but has not yet formed stars, or (2) a sign that the trend discussed above is not universal.

\section{Summary}
\label{sec:summary}

In this study we have mapped the {\CO} emission at $\sim40$~pc spatial and $\sim 1$~{\kms} spectral resolution in eleven star forming regions in the nearby spiral galaxy NGC~300 with the Submillimeter Array. We used the \texttt{CPROPS} algorithm with physically motivated priors to identify GMCs within our {\CO} data cubes and compute their masses, linewidths, and (where possible) sizes. We find that the CO emission seen as a single pixel in the APEX survey of \cite{Faesi:2014ib} is resolved into two or more discreet GMCs in each region. Our sample consists of 45 total GMCs, 12 of which are spatially resolved (the ``resolved subsample'').

\begin{enumerate}
\item{We extracted spectra from each GMC, and find that the majority of spectra are well-fit by a single Gaussian, suggesting that these GMCs are indeed individual clouds to the limit of our spectral resolution. Two of the resolved subsample appear to show evidence for multiple emission components and therefore we may have overestimated their actual line widths and virial masses in this analysis.}
\item{We detect $^{13}$CO in four of the most massive clouds in the sample, and in those clouds find an average {\CO}/$^{13}$CO flux ratio of 6.2.}
\item{We find that masses derived directly from CO luminosity and virial masses are in relative agreement for those GMCs without evidence for multiple spectral emission components. This suggests that the majority of the NGC~300 clouds are virialized, similarly to GMCs in the Milky Way and other nearby galaxies. The exception, DCL79-2, likely has a size underestimated by \texttt{CPROPS} and potentially a steep mass density profile.}
\item{We fit the GMC mass spectrum in this sample of NGC~300 clouds and derive a slope of $\gamma=1.8 \pm 0.07$. This value is intermediate between the slopes of the mass spectra in the inner ($\gamma=1.5$) and outer (2.1) Milky Way, and similar to the mass spectrum slopes in the inner regions of M33 (1.6) and in the spiral arms of M51 (1.8). Since our sample consists of GMCs near {\HII} regions and at galactocentric radii less than 4 kpc, this result suggests a similarity in mass spectrum slope in inner galaxy star-forming environments across galaxies.}
\item{The resolved subsample shows consistency with the \cite{Larson:1981vv} size-linewidth, mass-size, and mass-linewidth relations seen in the Milky Way and many other galaxies. We fit the resolved subsample data using orthogonal distance regression fitting and find a statistical significant trend of $\Delta V \propto R^{0.52\pm0.20}$. This result is in agreement with studies of other nearby galaxies, though our data preclude any conclusive statements regarding the origin and nature of the turbulence in GMCs. The surface densities of the resolved clouds range from 25 to 125~{\Msun}~pc$^{-2}$, with a median of 54~{\Msun}~pc$^{-2}$, similar to GMC surface densities in Milky Way clouds.}
\item{The fraction of the APEX single dish integrated intensity we recover with the SMA ranges from 30\% to nearly 100\% across our sample. Low detection fractions are likely due to being unable to detect low-mass GMCs with our limited sensitivity observations as compared with the single dish measurements, and also possibly the presence of some diffuse gas. The SMA/APEX integrated intensity ratio is not correlated with the APEX molecular mass or galactocentric radius. There is however a trend with the SMA integrated intensity, suggesting systematic variation in the contribution of massive GMCs to the total CO intensity of the region. In addition, for regions with star formation rates larger than $5 \times 10^{-4}~\Msun$~yr$^{-1}$, we tend to recover a higher fraction of the single dish flux in regions with high SFRs than in those with low SFRs. This result could be explained by the dissociation of low mass clouds and diffuse gas due to energetic ultraviolet radiation from the many newly formed massive stars in the high SFR regions.}
\end{enumerate}

To test these conclusions, additional high-resolution observations of a larger sample size of NGC~300 clouds will be needed. With the resolution and sensitivity of ALMA, for example, the entire GMC sample of $\sim50$ regions from \cite{Faesi:2014ib} could be studied at $\sim10$~pc resolution in a very reasonable amount of observing time. Such studies will be crucial to understanding the general properties and broad populations of GMCs.

The authors owe their deepest gratitude to Nimesh Patel for tireless guidance on the use of \texttt{MIR-IDL} and \texttt{Miriad} for SMA data reduction and imaging. We also wish to thank Erik Rosolowsky for helpful discussions on the application of \texttt{CPROPS} to this data set. We acknowledge the anonymous referee for extremely helpful comments that greatly improved the quality of this manuscript. In addition, we thank Eric Keto for assistance in obtaining the observations leading to this paper. C.M.F acknowledges support from a National Science Foundation Graduate Research Fellowship under Grant No. DGE-1144152.



\clearpage
\begin{thebibliography}{34}
\expandafter\ifx\csname natexlab\endcsname\relax\def\natexlab#1{#1}\fi

\bibitem[{Bertoldi \& McKee(1992)}]{1992ApJ...395..140B}
Bertoldi, F., \& McKee, C. F. 1992, ApJ, 395, 140


\bibitem[{Bolatto {et~al.}(2008)Bolatto, Leroy, Rosolowsky, Walter, \&
  Blitz}]{2008ApJ...686..948B}
Bolatto, A.~D., Leroy, A.~K., Rosolowsky, E., Walter, F., \& Blitz, L. 2008,
  ApJ, 686, 948

\bibitem[{Brinchmann {et~al.}(2004)Brinchmann, Charlot, White, Tremonti,
  Kauffmann, Heckman, \& Brinkmann}]{2004MNRAS.351.1151B}
Brinchmann, J., Charlot, S., White, S. D.~M., {et~al.} 2004, MNRAS, 351, 1151

\bibitem[{Colombo {et~al.}(2014)Colombo, Hughes, Schinnerer, Meidt, Leroy,
  Pety, Dobbs, Garc{\'\i}a-Burillo, Dumas, Thompson, Schuster, \&
  Kramer}]{2014ApJ...784....3C}
Colombo, D., Hughes, A., Schinnerer, E., {et~al.} 2014, ApJ, 784, 3

\bibitem[{Deharveng {et~al.}(1988)Deharveng, Caplan,
  Lequeux, Azzopardi, Breysacher, Tarenghi, \& Westerlund}]{Deharveng:1988wh}
Deharveng, L., Caplan, J., Lequeux, J., {et~al.} 1988, A\&AS, 73,
  407

\bibitem[{Donovan~Meyer {et~al.}(2013)Donovan~Meyer, Koda, Momose, Mooney,
  Egusa, Carty, Kennicutt, Kuno, Rebolledo, Sawada, Scoville, \&
  Wong}]{2013ApJ...772..107D}
Donovan~Meyer, J., Koda, J., Momose, R., {et~al.} 2013, ApJ, 772, 107

\bibitem[{Engargiola {et~al.}(2003)Engargiola, Plambeck, Rosolowsky, \&
  Blitz}]{2003ApJS..149..343E}
Engargiola, G., Plambeck, R.~L., Rosolowsky, E., \& Blitz, L. 2003, ApJS, 149,
  343

\bibitem[{Faesi(2016)Faesi 2016}]{Faesi:2016kz}
Faesi, C.~M. 2016, Replication Data for: ``Resolving Giant Molecular Clouds in NGC 300: A First Look with the Submillimeter Array'', v1.0, Harvard Dataverse, doi:10.7910/DVN/IN7FZS

\bibitem[{Faesi {et~al.}(2014)Faesi, Lada, Forbrich, Menten, \&
  Bouy}]{Faesi:2014ib}
Faesi, C.~M., Lada, C.~J., Forbrich, J., Menten, K.~M., \& Bouy, H. 2014, ApJ,
  789, 81

\bibitem[{Fukui \& Kawamura(2010)}]{Fukui:2010ki}
Fukui, Y., \& Kawamura, A. 2010, ARA\&A, 48, 547

\bibitem[{Fukui {et~al.}(2008)Fukui, Kawamura, Minamidani, Mizuno, Kanai,
  Mizuno, Onishi, Yonekura, Mizuno, Ogawa, \& Rubio}]{2008ApJS..178...56F}
Fukui, Y., Kawamura, A., Minamidani, T., {et~al.} 2008, ApJS, 178, 56

\bibitem[{Gieren {et~al.}(2004)Gieren, Pietrzy{\'n}ski, Walker, Bresolin,
  Minniti, Kudritzki, Udalski, Soszy{\'n}ski, Fouqu{\'e}, Storm, \&
  Bono}]{2004AJ....128.1167G}
Gieren, W., Pietrzy{\'n}ski, G., Walker, A., {et~al.} 2004, AJ, 128, 1167

\bibitem[{Gratier {et~al.}(2012)Gratier, Braine, Rodriguez-Fernandez, Schuster,
  Kramer, Corbelli, Combes, Brouillet, van~der Werf, \&
  R{\"o}llig}]{Gratier:2012km}
Gratier, P., Braine, J., Rodriguez-Fernandez, N. J., {et~al.} 2012, A{\&}A, 542, A108

\bibitem[{Heyer \& Dame(2015)}]{Heyer:2015ee}
Heyer, M., \& Dame, T.~M. 2015, ARA\&A, 53, 583

\bibitem[{Heyer {et~al.}(2009)Heyer, Krawczyk, Duval, \&
  Jackson}]{Heyer:2009ii}
Heyer, M., Krawczyk, C., Duval, J., \& Jackson, J.~M. 2009, ApJ, 699, 1092

\bibitem[{Heyer {et~al.}(2001)Heyer, Carpenter, \& Snell}]{2001ApJ...551..852H}
Heyer, M.~H., Carpenter, J.~M., \& Snell, R.~L. 2001, ApJ, 551, 852

\bibitem[{Inutsuka {et~al.}(2015)Inutsuka, Inoue, Iwasaki, \&
  Hosokawa}]{Inutsuka:2015gm}
Inutsuka, S.-i., Inoue, T., Iwasaki, K., \& Hosokawa, T. 2015, A{\&}A, 580, A49

\bibitem[{Jackson {et~al.}(2006)Jackson, Rathborne, Shah, Simon, Bania,
  Clemens, Chambers, Johnson, Dormody, Lavoie, \& Heyer}]{2006ApJS..163..145J}
Jackson, J.~M., Rathborne, J. M., Shah, R. Y., {et~al.} 2006, ApJS, 163, 145

\bibitem[{Kirk {et~al.}(2015)Kirk, Gear, Fritz, Smith, Ford, Baes, Bendo,
  de~Looze, Eales, Gentile, Gomez, Gordon, O'Halloran, Madden, Roman-Duval,
  Verstappen, Viaene, Boselli, Cooray, Lebouteiller, \&
  Spinoglio}]{2015ApJ...798...58K}
Kirk, J.~M., Gear, W. K., Fritz, J., {et~al.} 2015, ApJ, 798, 58

\bibitem[{Kritsuk {et~al.}(2013)Kritsuk, Lee, \& Norman}]{2013MNRAS.436.3247K}
Kritsuk, A.~G., Lee, C.~T., \& Norman, M.~L. 2013, MNRAS, 436, 3247

\bibitem[{Lada {et~al.}(1988)Lada, Margulis, Sofue, Nakai, \&
  Handa}]{1988ApJ...328..143L}
Lada, C.~J., Margulis, M., Sofue, Y., Nakai, N., \& Handa, T. 1988, ApJ, 328, 143

\bibitem[{Larson(1981)}]{Larson:1981vv}
Larson, R.~B. 1981, MNRAS, 194, 809

\bibitem[{Leroy {et~al.}(2009)Leroy, Walter, Bigiel, Usero, Weiss, Brinks,
  de~Blok, Kennicutt, Schuster, Kramer, Wiesemeyer, \& Roussel}]{Leroy:2009di}
Leroy, A.~K., Walter, F., Bigiel, F., {et~al.} 2009, AJ, 137, 4670

\bibitem[{Lombardi {et~al.}(2010)Lombardi, Alves, \&
  Lada}]{2010A&A...519L...7L}
Lombardi, M., Alves, J., \& Lada, C.~J. 2010, A{\&}A, 519, L7

\bibitem[{Rebolledo {et~al.}(2012)Rebolledo, Wong, Leroy, Koda, \&
  Meyer}]{Rebolledo:2012ex}
Rebolledo, D., Wong, T., Leroy, A., Koda, J., \& Meyer, J.~D. 2012, ApJ, 757,
  155

\bibitem[{Rieke {et~al.}(2004)Rieke, Young, Engelbracht, Kelly, Low, Haller,
  Beeman, Gordon, Stansberry, Misselt, Cadien, Morrison, Rivlis, Latter,
  Noriega-Crespo, Padgett, Stapelfeldt, Hines, Egami, Muzerolle,
  Alonso-Herrero, Blaylock, Dole, Hinz, Le~Floc'h, Papovich,
  P{\'e}rez-Gonz{\'a}lez, Smith, Su, Bennett, Frayer, Henderson, Lu, Masci,
  Pesenson, Rebull, Rho, Keene, Stolovy, Wachter, Wheaton, Werner, \&
  Richards}]{2004ApJS..154...25R}
Rieke, G.~H., Young, E.~T., Engelbracht, C.~W., {et~al.} 2004, ApJS, 154, 25

\bibitem[{Rodighiero {et~al.}(2011)Rodighiero, Daddi, Baronchelli, Cimatti,
  Renzini, Aussel, Popesso, Lutz, Andreani, Berta, Cava, Elbaz, Feltre,
  Fontana, F{\"o}rster~Schreiber, Franceschini, Genzel, Grazian, Gruppioni,
  Ilbert, Le~Floc'h, Magdis, Magliocchetti, Magnelli, Maiolino, McCracken,
  Nordon, Poglitsch, Santini, Pozzi, Riguccini, Tacconi, Wuyts, \&
  Zamorani}]{2011ApJ...739L..40R}
Rodighiero, G., Daddi, E., Baronchelli, I., {et~al.} 2011, ApJ, 739, L40

\bibitem[{Roman-Duval {et~al.}(2016)Roman-Duval, Heyer, Brunt, Clark, Klessen,
  \& Shetty}]{2016arXiv160100937R}
Roman-Duval, J., Heyer, M., Brunt, C., {et~al.}
  2016, arXiv e-prints, arXiv:1601.00937

\bibitem[{Rosolowsky(2005)}]{Rosolowsky:2005gt}
Rosolowsky, E. 2005, PASP, 117, 1403

\bibitem[{Rosolowsky {et~al.}(2003)Rosolowsky, Engargiola, Plambeck, \&
  Blitz}]{2003ApJ...599..258R}
Rosolowsky, E., Engargiola, G., Plambeck, R., \& Blitz, L. 2003, ApJ, 599, 258

\bibitem[{Rosolowsky {et~al.}(2007)Rosolowsky, Keto, Matsushita, \&
  Willner}]{2007ApJ...661..830R}
Rosolowsky, E., Keto, E., Matsushita, S., \& Willner, S.~P. 2007, ApJ, 661, 830

\bibitem[{Rosolowsky \& Leroy(2006)}]{Rosolowsky:2006cb}
Rosolowsky, E., \& Leroy, A. 2006, PASP, 118, 590

\bibitem[{Sakamoto {et~al.}(1997)Sakamoto, Hasegawa, Handa, Hayashi, \&
  Oka}]{1997ApJ...486..276S}
Sakamoto, S., Hasegawa, T., Handa, T., Hayashi, M., \& Oka, T. 1997,
  ApJ, 486, 276

\bibitem[{Sault {et al.}(1995)Sault, Teuben, \& Wright}]{1995ASPC...77..433S}
Sault, R.~J., Teuben, P.~J., \& Wright, M.~C.~H. 1995, adass, 77, 433

\bibitem[{Solomon {et~al.}(1987)Solomon, Rivolo, Barrett, \&
  Yahil}]{Solomon:1987uq}
Solomon, P.~M., Rivolo, A.~R., Barrett, J., \& Yahil, A. 1987, ApJ, 319, 730

\bibitem[{Solomon {et~al.}(1979)Solomon, Sanders, \&
  Scoville}]{1979ApJ...232L..89S}
Solomon, P.~M., Sanders, D.~B., \& Scoville, N.~Z. 1979, ApJ,
  232, L89

\bibitem[{Steer {et~al.}(1984)Steer, Dewdney, \& Ito}]{1984A&A...137..159S}
Steer, D.~G., Dewdney, P.~E., \& Ito, M.~R. 1984, A\&A, 137, 159

\bibitem[{Ulich \& Haas(1976)}]{1976ApJS...30..247U}
Ulich, B.~L., \& Haas, R.~W. 1976, ApJ, 30, 247

\end{thebibliography}
\end{document}